\documentclass[10pt]{article}
\usepackage{movie15}
\usepackage{wrapfig}
\usepackage{color}
\usepackage{graphicx}
\usepackage[colorlinks=true, pdfstartview=FitV, linkcolor=blue, citecolor=blue, urlcolor=blue]{hyperref}

\usepackage{amsbsy}
\usepackage{amssymb}

\usepackage{verbatim}
\makeatletter
\@addtoreset{equation}{section}
\makeatother

\newtheorem{theorem}{Theorem}[section]

\newtheorem{exercise}{Exercise}[section]

\newtheorem{lemma}{Lemma}[section]
\newtheorem{remark}{Remark}[section]

\newtheorem{proposition}{Proposition}[section] 
\newtheorem{corollary}{Corollary}[section] 
\newtheorem{definition}{Definition}[section]
\def\le{\left}
\def\ri{\right}
\def\ds{\displaystyle}
\def\V {\mathcal V}

\def\un{\underline}
\def\res{\mathop{\mathrm {res}}\limits_}

\def\br{\begin{remark}}
\def\er{\end{remark}}
\def\bt{\begin{theorem}}
\def\et{\end{theorem}}
\def\bc{\begin{corollary}}
\def\ec{\end{corollary}}
\def\bx{\begin{examp}\small}
\def\ex{\end{examp}}
\def\bxr{\begin{exercise}\small}
\def\exr{\end{exercise}}
\def\bl{\begin{lemma}}
\def\el{\end{lemma}}
\def\bd{\begin{definition}}
\def\ed{\end{definition}}
\def\bp{\begin{proposition}}
\def\ep{\end{proposition}}
\def\be{\begin{equation}}
\def\ee{\end{equation}}
\def\&{\hspace{-15pt}&}
\def\bea{\begin{eqnarray}}
\def\eea{\end{eqnarray}}
\def\beas{\begin{eqnarray*}}
\def\eeas{\end{eqnarray*}}
\def\B{\boldsymbol {\mathfrak B}}

\def\C{{\mathbb C}}
\def\L{\mathcal L}
\def\R{{\mathbb R}}
\def\N{{\mathbb N}}

\def\H{{\cal H}}

\def\a{\alpha}
\def\d{\,\mathrm d}

\def\Pot{\mathcal {P}}
\def\1{{\bf 1}}
\def\wt{\widetilde}

\date{}
\begin{document}
\baselineskip 15pt plus 1pt minus 1pt
\begin{titlepage}
\vspace{0.2cm}
\begin{center}
\begin{Large}
\fontfamily{cmss}
\fontsize{17pt}{27pt}
\selectfont
\textbf{Boutroux curves with external field: equilibrium measures without a minimization problem}
\end{Large}\\
\bigskip
\begin{large} {M.
Bertola}$^{\ddagger,\sharp}$\footnote{Work supported in part by the Natural
    Sciences and Engineering Research Council of Canada (NSERC).}\footnote{bertola@crm.umontreal.ca}
\end{large}
\\
\bigskip
\begin{small}
$^{\ddagger}$ {\em Department of Mathematics and
Statistics, Concordia University\\ 1455 de Maisonneuve W., Montr\'eal, Qu\'ebec,
Canada H3G 1M8} \\
$^{\sharp}$ {\em Centre de recherches math\'ematiques\\ Universit\'e\ de Montr\'eal } \\
\end{small}
\bigskip
{\bf Abstract}
\end{center}

The nonlinear steepest descent method for rank-two systems relies on the notion of $g$-function.
The applicability of the method ranges from orthogonal polynomials (and generalizations) to Painlev\'e\ transcendents, and integrable wave equations (KdV, NonLinear Schr\"odinger ,etc.).

 For the case of asymptotics of generalized orthogonal polynomials with respect to varying {\em complex} weights we can recast the requirements  for the Cauchy-transform of the equilibrium measure into a  problem of algebraic geometry and harmonic analysis and completely solve the existence and uniqueness issue without relying on the minimization of a functional. This addresses and solves also the issue of the ``free boundary problem'', determining implicitly the curves where the zeroes of the orthogonal polynomials accumulate in the limit of large degrees and the support of the measure. The relevance to the quasi---linear Stokes phenomenon for Painlev\'e\ equations is indicated.
A numerical algorithm to find these curves in some cases is also explained.

{\tiny Technical note: the animations included in the file can be viewed using Acrobat Reader 7 or higher. Mac users should also install  a QuickTime plugin called Flip4Mac. Linux users can extract the embedded animations and play them with an external program like VLC or MPlayer. All trademarks are owned by the respective companies.}
\vspace{0.7cm}
\begin{center}
\includegraphics[width=11cm]{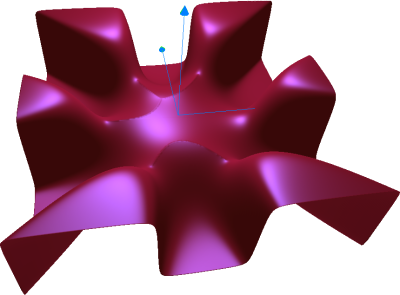}
\end{center}
\end{titlepage}

\begin{figure}
\caption{The surface of (the arctan of) $h(x)$ (see the text for explanation) for an admissible Boutroux simple curve with external potential $V(x) = x^6/6$. Note the ``creases'' where $h$ is clearly non-differentiable but continuous; on each side of each crease the surface is negative.}
\label{3Dgalore}
\end{figure}
\tableofcontents
\section{Introduction and setting}
In the  nonlinear steepest descent method \cite{Deift} the ``cornerstone'' is the construction of a suitable $g$--function for the problem: for systems of rank $2$ there is a single $g$--function whereas in general (disposing of a trivial gauge term) for a rank $(r+1)$--system one needs a  $r$-tuple of $g$--functions.

The rank $2$ case is sufficiently rich so as to include systems relevant to all Painlev\'e\ transcendents \cite{JMU}  as well orthogonal polynomials \cite{DKMVZ} and generalizations thereof to complex measures \cite {BertoMo, ItsKitaev} and/or to different notions of biorthogonality of Laurent polynomials \cite{BertoMisha}.

The present paper will not deal with the steepest--descent method itself but rather with the construction of suitable $g$--functions of the type of interest in the study of pseudo--orthogonal polynomials. Once the method has been established it is conceptually simple  (but possibly practically complicated) to extend the present considerations to other settings like the linear auxiliary systems appearing in Painlev\'e\ equations and Laurent--orthogonal polynomials.\par \vskip 6pt
\begin{wrapfigure}{r}{0.4\textwidth}
\resizebox{6cm}{!}{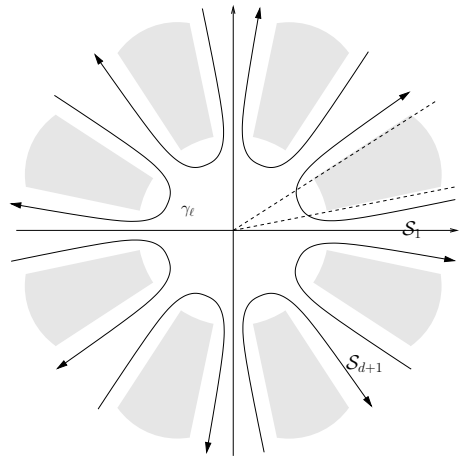}
\caption{An example of contours for a potential $V(x)$ of degree $8$. The shaded sectors are ``forbidden'' directions of approach at infinity. The contours $\gamma_\ell$ can approach $\infty$ along any direction in the ``allowed sectors'', non shaded and marked as $\mathcal S_1, \dots, \mathcal S_{d+1}$. We should think of the {\em oriented} contours $\gamma_\ell$ as a {\bf wire} or {\bf highway} carrying a current (traffic) $\varkappa_\ell\in \C$: the {\bf total incoming current} in a sector $\mathcal S_j$ is the sum of the currents in/out carried by all contours accessing that particular sector. Note that the total traffic in/out of all sectors is zero.}
\label{contours}
\end{wrapfigure}
The most direct motivation of the present work essentially stems from a previous paper \cite{BertoMo} dealing with pseudo--orthogonal polynomials described below: we point out that such pseudo--orthogonal polynomial were already studiend in \cite{ItsKitaev} in the context of quantum gravity.

Consider a {\bf potential} ({\bf external field}) $V(x)$: we will only focus on polynomials in the present paper but similar techniques can be used to handle a potential with arbitrary rational derivative (we are presently working on this extension) and certainly some ad-hoc situations of even more general form.

Let thus $V(x)$ be a polynomial of degree $d+1$ and  consider now  the complex {\bf moment functional} (a ``semiclassical'' moment functional in the language of \cite{marcellan, BEHsemi})
\bea
\L_{\varkappa,N}:&\& \C[x] \mapsto \C\\
	&\& x^j \mapsto \sum_{\ell =1}^{d+1} \varkappa_\ell \int_{\gamma_\ell} x^j{\rm e}^{-N V(x)} 
\label{momfunct}
\eea

where the complex constants $\varkappa_\ell$ are arbitrarily chosen but fixed and $N\in \R_+$ is a large parameter used in the asymptotic study below. The contours $\gamma_\ell$ are some contours in the complex plane extending to infinity along directions where $\Re (V(x))\to +\infty$, so as to give convergent integrals. Note that the details of the contours are irrelevant in the definition of the moment functional by virtue of Cauchy's theorem. Also, by the same argument, one of the contours is redundant in the sense that the integral along one of them can be expressed as minus the sum of the integral along all the others. We will nevertheless retain this ``redundant'' definition for later purposes.

The {\bf pseudo--orthogonal polynomials} $p_n(x)$ are a sequence of (monic) polynomials of exact degree $n$ with the defining properties \cite{BEHsemi}
\be
\L_{\varkappa,N}(p_n(x) p_m(x)) = h_{n} \delta_{nm}\ , \ \ h_n\neq 0\ ,\forall n\in \N.
\ee
The reader should recognize at this point the connection to ordinary OPs; indeed if $V(x)$ is real-valued and we choose only the real axis as contour, then the definition above reduces to the standard one.

The existence of these generalized OPs amounts to the non vanishing of the sequence of  H\"ankel determinants of the moment functional \cite{BEHsemi}. The problem is that of studying the asymptotics of the polynomials $p_n(x)$ in the complex plane uniformly over compact sets as $N\to \infty$ and $n/N\to T\in [0,\infty)$. 

This problem was addressed and solved\footnote{In \cite{BertoMo} we considered only the case $n = N+r$ with $r$ bounded, but it is a matter of a simple modification to extend that work to the case $n = T N+r$.} in \cite{BertoMo} using the nonlinear steepest descent method of Deift--Zhou  and hence a suitable notion of $g$--function. However we solved the problem  in a completely ``backwards'' perspective: indeed in that paper  the data was not a given potential $V$ but rather a given Boutroux hyperelliptic curve (satisfying certain conditions of ``admissibility'') and the potential itself was encoded in certain moduli. The relation between those moduli and the coefficients of the potential is highly transcendental (except for the degree of the potential which is immediately read off)  inasmuch as it involves the solution of a uniformization problem: thus the question of how to construct a suitable $g$--function starting from the potential $V$ remained unanswered.

We remind the reader that in the case of a real potential on the real axis the solution is provided by a variational problem for a functional 
\be
\mathcal S[\rho] := 2\int_\R V(x) \rho(x)\d x - \int_\R \int_\R \rho(x)\rho(x') \ln | x-x'|\ ,
\ee
where $\rho(x) \d x$ is a positive measure of total mass $T$. It is known that under mild assumptions (certainly satisfied for a polynomial $V$ bounded from below) the measure minimizing the above functional is unique and supported on a finite disjoint union of compact intervals \cite{SaffTotik, McLaughlinDeiftKriecherbauer}. The $g$--function is then defined as 
\be
g(x) = V(x) - \int_\R \ln(x-\xi) \rho(\xi)\d \xi
\ee
where the cuts of the logarithm are suitably placed on the real axis. 
It follows from a manipulation involving the Plemelji--Sokhotsky formula that the holomorphic derivative of $g(x)$ satisfies an algebraic equation
\be
y^2 = (V'(x))^2 -  Q(x)\ ,\label{hyper1}\ \ \ y:= g'(x)\ ,\ \ Q(x) = \int_{\R} \frac{V'(x)-V'(\xi)}{x-\xi} \rho(\xi)\d \xi
\ee
where --if $V(x)$ is a polynomial of degree $d+1$-- $Q(x)$ is a polynomial of degree $d-1$: of course $Q(x)$ is determined implicitly as the solution of the above variational problem.

In the case of a complex potential and (weighted) sum of contours other than the real axis, a similarly developed variational problem is not known, although it is believed that the minimization problem should be replaced by a ``min-max'' problem where one minimizes a similar functional on the measures and then maximizes it over the variation of the contours within a carefully determined class. The obstacle with this approach from the analytical point of view is that it introduces a ``free--boundary problem'' which is in general hard to pin-down also because the smoothness class of the boundary is not known a priori.\footnote{There is one work \cite{Rakhmanov} which deals with a problem that falls within the same circle of investigations where the authors use the notion of $S$--curve proving its existence: we point out that the present method is radically different inasmuch as it relies  on  ``differential geometry''  rather than analysis.}

The point of interest is that the variational conditions can be translated into properties of the algebraic equation (\ref{hyper1}) in a way that allows straightforward generalization without explicit reference to a minimization problem at all. This re-formulation is contained in the notion of {\bf Boutroux curve} and {\bf admissibility} introduced in \cite{BertoMo} and recalled in the next section.

For the reader with experience in the nonlinear steepest--descent method we point out  that these requirements are the sufficient conditions for fully implementing the Deift--Zhou analysis and thus bypassing entirely the solution of a minimization problem (as explained in \cite{BertoMo}) although being logically equivalent to it\footnote{In fact a minimization problem is conceptually needed when solving a certain Dirichlet problem for a uniformizing map.}.

\subsection{Main results of the paper}
The main result of the paper is Theorem \ref{maintheorem}; in coarse terms it asserts that for any (polynomial) potential $V(x)$ and total charge $T>0$ there exist (unique) an appropriate $g$--function for each choice of contours in the moment functional (\ref{momfunct}) which is suitable for the nonlinear steepest descent analysis. The knowledge of such $g$--function is equivalent to the knowledge of a positive measure (the {\bf equilibrium measure}) supported on certain arcs in the complex plane (Sect. \ref{eqmeasure}), and the contours appearing in (\ref{momfunct}) can be smoothly deformed (within the same ``homology'' class) so that they pass through these arcs and is such a way that all the inequalities required by the steepest descent method are fullfilled (this was explained in \cite{BertoMo}).

The idea of the proof is rather differential--geometric and it is based on  two ingredients:
\begin{itemize}
\item proving that there is some $g$--function for a {\em similar} problem which has the same contours as in (\ref{momfunct}) but for a potential $\wt V(x)$ (polynomial of the same degree as $V(x)$) whose coefficients are  not explicitly known;
\item proving that we can deform the underlying hyperelliptic curve as in eq(\ref{hyper1}) so that the ``connectivity''\footnote{The precise definition of this word will appear later in Sect. \ref{secfive}.} is preserved and the potential can be made to match any given one (Sects. \ref{secdeform} and following).
\end{itemize}
The precise statements are contained only in Sect. \ref{secdeform} and following,  because  they require introduction of a number of concepts. 
Indeed in Section \ref{secboutroux} we review the notions of {\em Boutroux curves} and  {\em admissibility}; in Sect. \ref{sectConstr} we recall the construction of Boutroux curves contained in \cite{BertoMo} and based on Strebel's theory of quadratic differentials \cite{Strebel}. This section also contains important {\em terminology} that is liberally used throughout the rest of the paper.

The map $\Pot$ that associates to a Boutroux curve its potential (defined in Def. \ref{extpotential}) is one of the crucial objects; the important point is that this is local isomorphism of spaces (Boutroux curves vs. potential/charge) but it is  a {\bf branched map}; Sect. \ref{secdim}, although still  preparatory, contains the necessary analysis of the local isomorphism Lemma \ref{lemmalift} and the study of the branch-locus (Sect. \ref{discriminantgeometry}).

Each sheet of $\Pot$ (a {\bf cell}) is attached to another sheet along the ramification locus; the main idea of the deformation argument is that we can lift a path in the potentiall/charge space to the space of Boutroux curves while retaining all the ``connectivity'' and admissibility properties even if we cross the branchlocus. This involves a gluing (Sect. \ref{sectsing}) of the coordinate chart given by $\Pot$ in each cell along the boundary consisting of the ramification locus. 
Finally in Sect. \ref{secunique} the uniqueness of the $g$--function is established.

Some indication of other applications (nonlinear Stokes phenomenon and other settings) is pointed out in the conclusion. In the appendices we provide some examples and  it is explained how to obtain numerically  Boutroux curves for a given potential.

\section{Boutroux curves}
\label{secboutroux}
We will consider only polynomial (as opposed to  rational) hyperelliptic Boutroux curves of the form 
\be
y^2 = P(x) \label{defcurve}
\ee 
where $P(x)$ is a polynomial of even degree: the case of polynomials of odd degrees would require trivial (but substantial) modifications and we prefer not to touch the issue here since the curve relevant for pseudo--orthogonal polynomial is anyways always of even degree\footnote{There are other situations, notably Painlev\'e\ I, where the curve would be of degree $3$, hence either rational or elliptic.}. The roots of $P(x)$ need not be distinct and hence the hyperelliptic curve may be nodal\footnote{As long as the genus of the surface of $y$ is finite one may easily consider more general analytic (entire) functions $P(x)$ on the right hand side.}.

\bd
An algebraic curve defined as in (\ref{defcurve}) is said to satisfy the {\bf Boutroux condition} if 
\be
\oint_\gamma y\d x \in i\R
\ee 
for any closed loop on the Riemann surface of the algebraic function $y$.
The set of Boutroux curves (of fixed degree) will be denoted by $\B$ (or $\B_n$ if we want to make explicit the degree $n$ of $P(x)$). 
\ed
Note that the set of (polynomial) Boutroux curves is invariant under the affine group 
\be
x\mapsto \lambda x + b\ ,\ \ y \mapsto \frac y \lambda\ ,\ \ \ (\lambda, b) \in \C^\times\times \C\ .\label{GaugeGroup}
\ee

We denote by $\mathbf P_{n}$ the (vector) space of polynomials $P(x)$ of degree $n$; we can view the set of Boutroux curves  equivalently as the set of those polynomials $P(x)$ such that $\Re \oint_\gamma \sqrt{P(x)}\d x =0$ for all closed contours $\gamma$ such that the sum of the multiplicities of the enclosed zeroes of $P(x)$  is even. Clearly the loops encircling only even--multiplicity roots give trivial constraints.

In the rest of the paper we will only consider polynomial Boutroux curves of degree $2d$ and we will hence omit any explicit reference to the degree when using the symbol $\B$. 
Let $2g+2$ be the number of odd--multiplicity roots $\{\alpha_j\}$ of $P(x)$ (each counted discarding multiplicity): note that necessarily there are an even number of them.  We will denote by 
\be
w^2 = \prod_{j=1}^{2g+2}(x-\alpha_j) \label{smoothcurve}
\ee
the smooth hyperelliptic curve with those branch-points. We will regard $y$ as a meromorphic function on the genus-$g$ hyperelliptic Riemann surface of $w$, where 
\be
y = M(x) w
\ee
for a suitable polynomial $M(x)$.

\bd
\label{potential}
For a given Boutroux curve $y^2 = P(x)$ we introduce the {\bf external potential (field)} $V(z)$ by the formula
\be
V(z) = \res{\infty} \ln\le(1 - \frac {z}x\ri) y\d x\ ,
\ee
and by the formula
\be
T = \res{\infty} y\d x\ ,
\ee
we define the {\bf total charge}, which is a real number because of the Boutroux condition.
The branch of $y$ is chosen so that it is the same in both formul\ae. The total charge is a real number (as follows from the Boutroux condition).  (Clearly the map $(V,T)\mapsto(-V,-T)$ corresponds to the exchange of sheets in the definition). 
\ed

\bd
\label{extpotential}
We will denote by $\V$ the set of pairs  $(V(x),T)$; this is naturally a manifold isomorphic to $\C^{d+1} \times \R$ (or $\R^{2d+3}$), the isomorphism being given by the $d+1$ coefficients of $V$ (recall that $V$ is constant-free) and the total charge $T$.  
The above defines a map 
\bea
\Pot:&\& \B  \mapsto \V \simeq   \C^{d+1}\times \R \simeq \R^{2d+3}\cr
&\& y\mapsto \Pot[y] = \le(V(x),T\ri) 
\eea
where $V(x)$ and $T$ are defined by the previous residue formul\ae\ in Def. \ref{potential}.

\ed
The branching locus of the map $\Pot$ will be studied in some detail later in Section \ref{secdim}.



Equivalently we may characterize the external field and the total charge by the asymptotic relation at infinity of the chosen branch of $y$
\be
y(x)  = \sqrt{P(x)} = V'(x) - \frac T x + \mathcal O(x^{-2}).
\ee
For a given Boutroux curve we define the {\bf admissible sectors} as the directions along which the real part of $V(x)$ tends to $+\infty$
\be
\mathcal S  = \{\theta\in [0,2\pi)\ : \ \ \Re(V(r{\rm e}^{i\theta})) \to + \infty, \ r\to +\infty\} = \bigsqcup_{j=1}^{d+1} \mathcal S_j\ ,\ \ d+1  := \deg V
\ee
In case of a potential with positive real leading coefficient (we can always assume this because of the action (\ref{GaugeGroup}), the admissible sectors and their complementary sectors (forbidden)  form a regular $2(d+1)$-gon, one side containing the positive real direction.
\bd
\label{defsimple}
A Boutroux curve is said to be
\begin{itemize}
\item  {\bf simple} if all roots are either simple or double;
\item {\bf noncritical} if for each root $m$ of multiplicity $\geq 2$ we have $\Re \int_{\alpha_1}^m y \d x\neq 0$, where $\alpha_1$ is a simple root\footnote{The integral is defined up to an overall sign but does not depend on the choice of the simple root $\alpha_1$ because of the Boutroux condition.}
\end{itemize}
This defines also the notion of {\bf critical} Boutroux curve.
We will denote the subset of simple and  noncritical curves by $\B_{reg}$, and note that it is an open subset of $\B$ (in the topology inherited from $\mathbf P$).
\ed 

\br
According to this definition, a Boutroux curve may be non-simple but also non-critical, as long as the root of multiplicity higher than $2$ is not on the zero levelset of the integral.
\er

\subsection{Admissible Boutroux curves}
In this section we recall the definition of {\em admissibility} as introduced in \cite{BertoMo}  and consequent properties.

From the Boutroux condition it follows that the locally defined harmonic function $h(z) = \Re\int_{\alpha_1}^z y\d x$ has only a multiplicative multivaluedness: harmonic continuation around any closed loop in $\C$ yields the same function up to a sign. In fact the sign is exactly the same obtained by analytic continuation of the algebraic function $y$. 
Because of the Boutroux condition the choice of basepoint $\a_1$ amongst the branchpoints is irrelevant: a different choice adds a half period --hence purely imaginary-- to the integral, leaving $h$ unaffected. Therefore the zero level set $\mathfrak X$  of $h(x)$ is intrinsically well defined and \cite{BertoMo} 
\begin{itemize}
\item the set $\mathfrak X$ consists of a finite union of Jordan arcs, some extending to $\infty$;
\item all branch-points $\alpha_j$ belong to $\mathfrak X$;
\item the set $\mathfrak X$ is topologically a forest of trees, namely it does not contain any loop;
\item each node is a zero of $P(x)         $ and has valency $k+2$, where $k$ is the multiplicity of the zero. In particular all the branchpoints are odd-valent nodes and viceversa all odd--valent nodes are branchpoints.
\end{itemize}

The definition of the external potential $V(z)$ and charge $T$ and the relationship with $y$ implies that one branch of $h(z)$ behaves as 
\be
h(z) \sim \Re V(z) - T \ln |z| + \mathcal O(1).\label{asymh}
\ee 

\bd
A Boutroux curve $y^2 = P(x)$ with total charge $T$ and external potential $V(x)$ is said to be {\bf pre-admissible} if the branch of $h(z)$ that behaves as in (\ref{asymh}) can be harmonically continued in a punctured neighborhood of $z=\infty$ and from there to a {\un{\em continuous}} function on  the \un{\em whole} complex plane $\C$. 
The set of preadmissible curves will be denoted by $\B^{<adm}\subset \B \subset \mathbf P$ and within it the simple--noncritical preadmissible by $\B_{reg}^{<adm}$.
\ed
There are a few important and almost immediate consequences of this definition which can be found in \cite{BertoMo}:
\bp
If the Boutroux curve is pre-admissible and noncritical (and hence all branchpoints are simple zeroes) then
\begin{enumerate}
\item The set $\mathcal B$ where $h$ is not harmonic is contained in the zero-level set $\mathfrak X$ and it is bounded; it will be called the {\bf branchcut structure} of the pre-admissible Boutroux curve.
\item The branchcut structure $\mathcal B$ consists of a finite union of finite Jordan arcs connecting the branchpoints.
\item Each connected component of $\mathcal B$ contains an even number of branchpoints.
\item Each arc of $\mathcal B$ is a {\bf critical trajectory} of the quadratic differential $P(x) \d x^2$ \cite{Strebel} on the complex plane, namely it is defined by the local ODE
\be
\nabla h(x) \dot x = \Re (y\d x) =0\ .
\ee
\end{enumerate}
\ep
\br
 The branchcut structure $\mathcal B$ is obtained by a process of pruning from the embedded graph $\mathfrak X$ as follows:\\
\parbox{12cm}{
\baselineskip 15pt plus 1pt minus 1pt
\begin{itemize}
 \item one removes from $\mathfrak X$ all the smooth branches, namely the maximal level curves $h(x)=0$ that do not contain a branchpoint in their closure. The reduced graph $\mathfrak X_0$ obtained this way consists only of critical trajectories issuing from one of the  branchpoints, each of which has valency $3$.
\item From $\mathfrak X_0$ we remove all the ``open branches'', namely the arcs that extend to infinity; this leaves nodes of our forest with valency one, two or three. 
\item  Each bivalent node belongs to a maximal chain of bivalent nodes; links in this chain are removed in the unique way that leaves nodes of valency $1$ along the chain  and so that all newly created connected components have an even number of nodes. The process is repeated to exhaustion until only mono and tri-valent nodes are left. 
\end{itemize}}\parbox{5cm}{\centerline{ \hspace{0.6cm}\includegraphics[scale=0.6]{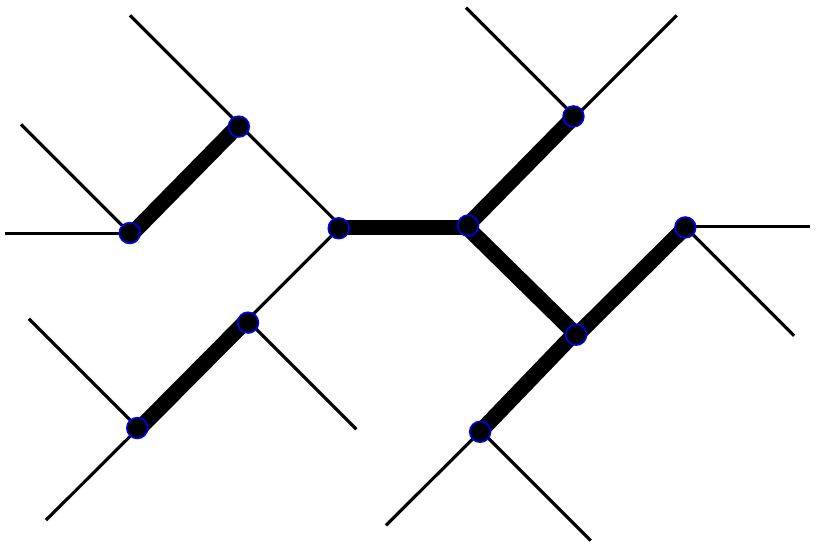}}}

A similar pruning (more complicated to describe) produces a suitable branchcut structure for an arbitrary pre-admissible Boutroux curve. In this more general setting one allows nodes of any valency greater than $3$ and start pruning in such a way that at the end  
\begin{itemize}
\item the parity of the valency of each node is unchanged;
\item the sum of the valency of the nodes in each connected component is even.
\end{itemize} 
The uniqueness of the result of such pruning is due to the absence of loops in the graph.
\er

Note that for a pre-admissible Boutroux curve with branchcut structure $\mathcal B$, each arc $\Gamma_k\in \mathcal B$ will be a critical trajectory joining two zeroes of $P(x)$. Because $h(z)$ is continuous but not harmonic on $\Gamma_k$ it must have the same sign on both sides of the arc, namely we have the 
\bl
For each smooth maximal arc $\Gamma\subset \mathcal B \setminus \{x:\ P(x)=0\}$ $h(\Gamma)=0$ and the sign of $h$ on the left and right of $\Gamma$ is the same. 
\el

This allows us to finally introduce the notion of admissibility.

\bd
A pre-admissible Boutroux curve $y^2 = P(x)$ with external potential $V(x)$, total charge $T>0$ and branchcut structure $\mathcal B$ is {\bf admissible} if $h$ is {\em negative} on both sides of each smooth maximal arc in $\mathcal B\setminus \{x:\ P(x)=0\}$.

The set of admissible Boutroux curves (and correspondingly the simple-noncritical admissible ones) will be denoted by  $\B^{adm}, \B^{adm}_{reg}$ and we have the following inclusions
\bea
\begin{array}{ccccccc}
\B^{adm} &\subset &\B^{<adm} &\subset &\B& \subset& \mathbf P\cr
\cup & & \cup && \cup && ||\cr
\B^{adm}_{reg} &\subset& \B^{<adm}_{reg} &\subset& \B_{reg} &\subset & \mathbf P
\end{array}
 .\label{sets}
\eea
Each of these sets inherits the natural topology from that of $\mathbf P$.
\ed

We point out that this notion of admissibility stems from the  specific application to the Deift-Zhou steepest descent method for the case of (pseudo)--orthogonal polynomials as explained in \cite{BertoMo}: in this setting the sign of $h$  being negative on the sides of each cut is necessary because the Stokes matrices are upper-triangular. The process of ``opening lenses'' \cite{DKMVZ} will give exponentially vanishing jumps in the approximate Riemann--Hilbert problem for the pseudo-orthogonal polynomials \cite{BertoMo}.
It is clear that if the applications requires elementary Stokes matrices of different triangularity, the notion of admissibility (and possibly the one of pre-admissibility) should be modified accordingly.

We conclude the section with the 
\bd [\cite{BertoMo}]
Given an admissible Boutroux curve with  branchcut structure  $\mathcal B$ the $g$--function is the locally analytic function 
\be
g(x) = \int_{\a_0}^x y\d x
\ee
defined on $\C\setminus \Gamma$, where $\Gamma $ is a collection of arcs containing the branchcuts and some additional arcs (arbitrarily chosen) connecting the components of $\mathcal B$ into a chain and to $\infty$, so as to have  a simply connected complement.
\ed
The main point that was raised in \cite{BertoMo} is that the notion of admissibility for the Boutroux curve is the set of  necessary and sufficient requirements for the $g$-function in order to apply the nonlinear steepest descent method.
\subsubsection{The (signed) measure associated to a (pre)admissible Boutroux curve}
\label{eqmeasure}
If $[y]\in \B^{<adm}$ then the jump of the normal derivative of $h$ across the branchcuts defines a real  measure; equivalently we can define the measure by 
\be
\frac 1{2i\pi}(y_+\d x - y_-\d x)  = \rho(x) \d |x|\ .
\ee
Such measure is real-analytic on the cuts: on each cut it has a definite sign and it is 
\begin{itemize}
\item {\bf positive} if the value of $h$ is negative on both sides of the cut;
\item {\bf negative} if the value of $h$ is positive on both sides of the cut.
\end{itemize}
In addition the total mass of the measure is precisely the total charge $T$ of the Boutroux curve. We leave these straightforward checks to the reader. We only point out (without proof) that 
\bp
There are no admissible Boutroux curves of negative total charge and  the only {\em admissible} Boutroux curves of zero charge are perfect squares $y^2 = (V'(x))^2$.
\ep
There are --however-- preadmissible Boutroux curves of negative charge and nontrivial ones of zero charge (but the corresponding measure is necessarily signed).

\section{Construction of Boutroux curves from combinatorial data and metric gluing}
\label{sectConstr}
In this section we recall the construction already expounded in \cite{BertoMo}, which allows to construct (admissible) Boutroux curves with prescribed topology of the branchcut structure.

In fact the statements do not have anything specific to Boutroux curves but are rather a convenient parametrization of the space of polynomials $P(x)$ in terms of certain combinatorial and metric data: we will not give the general account here for brevity and refer ibidem for further details.

The observation that drives the construction is that the Riemannian metric 
\be
\d s^2 = |P(x)| |\d x|^2
\ee
(here $|\d x|$ stands for the usual distance in the plane) is a {\bf flat metric} with conical singularities at the zeroes of $P(x)$; the flat coordinates of this metric are the real and imaginary parts of $G:= \int \sqrt {P(x)}\d x$, which are locally defined away from the zeroes of $P$. 

As in \cite{Strebel} we will use the notion of vertical\footnote{Here we deviate from the common use in the pertinent literature, which is that of horizontal trajectories. The difference is of pure convention and it is forced since we are interested in the level curves of $h(z) = \Re \int y\d x$, better described as ``vertical''.} critical trajectories: these are the maximal solutions of $\Re (y\d x)=0$ issuing from all the conical points (the zeroes). It is known (and a local computation shows it easily) that from a zero of multiplicity $k$ there are $k+2$ critical trajectories. 

We denote by 
\be
\Xi :=\{\hbox {closure of all maximal critical trajectories}\}
\ee
 the closure of the union  of all these trajectories: note that in our situation $\Xi$ includes also the critical trajectories issuing from the double (or higher order) zeroes of $P(x)$, which are saddles (or stationary points) of $h$.  

{\bf Terminology}. Suppose we are given an admissible Boutroux curve and have defined the harmonic function $h(z) = \Re \int_{\alpha_1}^z y\d x$ (as described earlier)  in such a way that it is continuous in $\C$, harmonic around $\infty$ and away from the (bounded) cuts and negative on the two sides of each cuts.

It is visually appealing and helpful for the intuition to think of the set $h^{-1}(\R_-)$ as the {\bf sea} and to the set $h^{-1}(\R_+)$ as the {\bf emerged lands} or {\bf continents}. The set $\mathfrak X = h^{-1}(0)$ consists of the branchcuts, which are surrounded by waters on the two sides, and non branchcuts, which have waters only on one side and will be referred as {\bf shorelines}. The branchcuts can be thought effectively as {\bf causeways} joining emerged continents.

Recall that \cite{Strebel, Spencer, BertoMo}
\begin{enumerate}
\item the complement $\C \setminus \Xi =\bigsqcup \Gamma_j$ is a finite union of {\bf unbounded,  simply connected regions} not containing any zero of $P(x)$ (which belong to $\Xi$ {\em a fortiori}).
\item Each $\Gamma_j$ has either one or two boundary components and hence is conformally a strip or a plane.
\item Each $\Gamma_j$ is uniformized by $W:= \int y \d x$ into a vertical strip (if it has two boundary components) or a halfplane (one boundary component).
\item On each boundary component of each $\Gamma_j$ there is at least one critical point (zero of $P(x)$). The relative positions in the $G$--plane of the uniformization of these marked points consists of the metric (moduli) data. For each plane they are the purely imaginary differences of the various marked points along the boundary; for strips there is also a complex number indicating the width/shear of the strip, relative to two arbitrarily chosen marked points on the opposite boundaries.
\item Each half plane $\Gamma_j$ in the $x$--plane contains all directions in a sector of the form $\theta \in \le( \frac {(2k-1)\pi}{2d+2}, \frac{(2k+1)\pi}{2d+2}\ri)$\footnote{We are assuming the Boutroux curve to be monic $P(x)= x^{2d} + \dots$. Otherwise all the sectors should be rotated by a common angle.}. There are thus precisely $2d+2$ halfplanes.
\item The two top rims of each strip go to infinity along the same direction, and so do the two bottom ones. Such directions are necessarily amongst the ``Stokes' lines'' $\theta = \frac {(2k+1)\pi}{2d+2}$, $k=0,\dots, 2d+1$.
\end{enumerate} 

{\bf Viceversa}, if we are given $2d+2$ halfplanes and a suitable number of strips together with marked points on the boundaries and a recipe detailing which strip/plane needs to be attached to which other, one can reconstruct the polynomial $P(x)$ by a process known as ``welding''  (a form of conformal gluing with prescribed flat metric) \cite{Strebel}.

In order to describe the topology of the gluing it was introduced in \cite{BertoMo} the notion of a ``clock diagram''. This is simply a $(2d+2)$--gon whose inside represents an abstract simply connected domain and whose sides represent asymptotic directions at infinity. The vertices of the polygon are then connected by a  suitable network  and the result is a  simplified visualization of the graph of critical lines $\Xi$.

In general a clock-diagram must have  precisely $2d+2$ (topological) halfplanes, each bordering one side of the $(2d+2)$-gon (i.e. a sector at $\infty$). The other domains (if any) must be topological strips.

By {\bf decorated} clock-diagram we mean the additional data of some positive numbers associated to the (undirected) links of the network between two nodes together with one modulus for each strip. The modulus of a strip is subordinated to the (arbitrary) choice of one node  on each of the opposite boundaries and is a complex number of nonzero real part. 

The meaning of the decorations is that of length of the critical trajectories between critical points in the flat Strebel metric. The moduli of the strips are the width (real part) and shear (imaginary part) of the chosen strip relative to the chosen reference marked points.
 
 The reconstruction theorem contained in \cite{BertoMo} asserts that
 \bt[\cite{BertoMo}]
 For an arbitrary decorated clock--diagram there exists  a polynomial $P(x)$ such that the critical graph $\Xi$ of the vertical trajectories has the same topology of the diagram and the same decorations (after suitable identification of the nodes).
 \et
We only recall the main idea and steps of the proof: it consists in gluing (metrically) a suitable number of flat metric halfplanes and strips according to the topology of the clock diagram: the decorations of the diagram are distances of marked points on the boundaries used as reference in the gluing.  

The result is  a topological (open) simply connected surface which is then compactified with one point to obtain a compact simply connected topological surface $X$.

Next one needs to define a conformal structure on $X$: near a point of the interior of the half-planes/strips, the conformal structure is defined using the flat coordinate $W$ itself. Near a conical point (where more than two regions are glued together) the conformal structure if defined by taking a suitable root of the flat coordinates $W$ in each of the regions. Some further care needs to be paid to define the conformal structure near the compactification point.

The result is a simply connected Riemann--surface  of genus zero.

If we denote by $W$ the flat (complex) coordinate in each of the pieces used in the gluing, one realizes that $W$ itself is not globally defined: moreover neither is $\d W$ since in the gluing it is necessary to flip some planes $W\to -W$. However $\d W^2$ is an invariantly and globally defined quadratic differential. In the conformal structure defined above it has  a number of zeroes at the nodes and a pole of suitable degree $2d+4$ at the compactification point.
Therefore in a uniformizing coordinate $x$, $\d W^2 = P(x) \d x ^2$ for some polynomial $P(x)$ of degree $2d$; the polynomial is defined only up to the group action of translations/dilations in $x$ (i.e. action \ref{GaugeGroup})  .\par\vskip 6pt
We point out that there is a quite transcendental step in going from the decorated clock--diagram to (some) polynomial $P(x)$, since the theorem involves the {\em uniformization theorem for Riemann surfaces}. This step involves ultimately the solution of a Dirichlet problem.

The Boutroux condition forces some constraints on the decoration and the topology of $\Xi$; however these constraints are linear in the decorations and hence pose no serious complication \cite{BertoMo}. 
We consider only the case of simple noncritical admissible Boutroux curves.
\begin{enumerate}
\item Given an emerged continent $C$ (connected) then the number of connected shorelines is precisely equal to the number of saddle-points in $C$  plus one, as follows from elementary Morse theory counting the components of $h^{-1}(r)$ as $r\searrow 0$.
\label{morse}
\item In a similar way (counting the the components of $h^{-1}(r), r \to +\infty$) each continent contains $k+1$ allowed sectors $\mathcal S_j$, where $k$ is the number of saddle points.
\item Similarly, the number of connected components of $\mathfrak X$ on the boundary of any ocean is equal to the number of saddle points in that ocean plus one.
\item A path connecting two components of $\mathfrak X_0$ crosses at least two strips. The sums of the width of the traversed strips (with appropriate signs that would be cumbersome to describe here) is zero.
\end{enumerate}
A lengthier and more detailed discussion can be found in \cite{BertoMo}.

\section {The map $\Pot$ and its branching: the discriminant $\bf \Sigma$}
\label{secdim}
We claim that $\Pot: \B\to \V$ (Def. \ref{extpotential}) is a branched covering: indeed the map $\Pot$ defines the coefficients of $V(x)$ and $T$ as polynomials in the coefficients of $P(x)$ (and rational in the square-root of the leading coefficient of $P(x)$). However this map is restricted to the sublocus $\B\subset \mathbf P$ and hence the branching occurs at certain type of Boutroux curves.


\begin{wrapfigure}{r}{0.37\textwidth}
\vspace{0.0cm}\resizebox{6cm}{!}{\includegraphics{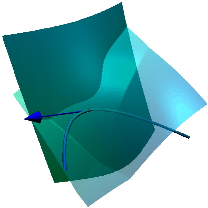}}
\vspace{-0.6cm}
\caption{A pictorial representation of the surface of the discriminant $\bf\Sigma$ in the $(V,T)$--space: it has singularities corresponding to the subloci where the Boutroux curves become more and more degenerate as well as self-intersections. Our path is chosen to  intersect it in the ``smooth''  part corresponding to the simple but (simply) critical Boutroux curves and transversally (i.e. the tangent to the path is not tangent to the discriminant). In the picture, the arc represents a path in the $\V$ space, and indicated is the tangent vector at the crossing of the discriminant. Note that the picutre is three--dimensional, but actual situation are always of higher dimension, so that this is only a suggestive picture.}
\label{FigDiscrim}
\vspace{-1cm}
\end{wrapfigure}

$\mathbf{Claim}$:
 A point $[y]\in \B$ is a branchpoint of the map $\Pot$ if and only if  the curve is critical   (Def. \ref{defsimple}), namely it has a root of multiplicity $\geq 2$ on the zero levelset $\mathfrak X= h^{-1}(0)$.\par \vskip 5pt 
We will not prove this in full generality; in this paper we will only consider {\em admissible} curves and  the most generic case of a double root on $\mathfrak X$. Indeed the content of Section \ref{sectsimple} will imply that in a neighborhood of a {\em critical admissible} Boutroux curve there are $3$ or $4$ (depending on the cases) preimages of $\Pot$.

The visual intuition is that  if a root $m$ of multiplicity $k\geq 2$ belongs to a shoreline or causeway, we can find nearby $[y]\in \B$ Boutroux curves with the same potential and total charge for which the root $m$ has either split in $k$ simple roots or  a collection of simple and double (noncritical) roots of total multiplicity $k$. For $k=2$ there are only two possibilities, namely the case where the double root $m$ has split into two simple roots (hence branchpoints and necessarily belonging to $\mathfrak X$) or else it  remains double but non-critical. From the combinatorial description  in terms of decorated clock--diagrams the reader should realize that  
\begin{itemize}
\item in order for a (single) double root to belong to $\mathfrak X$ the width of a strip (or the length of an edge) must tend to zero.
\end{itemize}

Therefore this is a ``(real) codimension-one'' occurrence. Any other way of approaching a critical Boutroux curve is of ``higher codimension''.
This prompts the following definition.
\bd
\label{defDiscriminant}
In the space of Boutroux curves we denote by $\mathbf \Delta \subset \B $ the {\bf branch-locus}, namely the co-dimension $1$ locus  of Boutroux curves that  are critical.

We denote by ${\mathbf \Delta}_{reg}\subset{\mathbf \Delta}$ the locus of critical but simple  Boutroux curves; inside it we denote by ${\mathbf \Delta}_{reg}^{0}$ the locus of {\bf simply critical} simple Boutroux curves, namely those for which only one saddle point belongs to the levelset $\mathfrak X = h^{-1}(0)$.

We denote by ${\mathbf \Delta}_{sing} \subset \mathbf \Delta$ the locus of non-simple, critical,  Boutroux curves.
The images under the map $\Pot$ of the above sets will be denoted by ${\mathbf \Sigma}^0_{reg}\subset{\mathbf \Sigma}_{reg}\subset {\mathbf \Sigma}\subset \V \simeq \R^{2d+3}$ and called the {\bf discriminant locus}.
\ed
\bd
\label{cell}
Each connected component of $\B_{reg}\setminus \mathbf \Delta$ (or $\B_{reg}^{<adm} \setminus \mathbf \Delta$ or  $\B_{reg}^{adm} \setminus \mathbf \Delta$ depending on the context) will be called a {\bf cell}.
\ed
The above discussion means that while ${\mathbf \Sigma}$ ($\mathbf \Delta$) chops the space $\V$ ($\B$) of potential/charges into possibly disconnected sets, the generic point in ${\mathbf \Sigma}$  belongs to ${\mathbf \Sigma}_{reg}^0$ ($\mathbf \Delta_{reg}^0$) and in a small neighborhood of that point ${\mathbf \Sigma}_{reg}^0$  ($\mathbf \Delta_{reg}^0$)  is a smooth manifold with the coordinates given by the Strebel decorations of the corresponding clock diagram. If the point belongs to the intersection of two or more smooth components of ${\mathbf \Sigma}_{reg}^0$, then the corresponding Boutroux curve will be simple and {\bf multiply critical}, namely with more than one double zero on the critical levelset: we can avoid this occurrence as well in general (but including this case would not really cause any serious obstacle).

Viceversa ${\mathbf \Sigma}_{sing}$ is of co-dimension two in the whole space and hence removing it (as well as multiply critical curves) leaves a connected space, where we can connect any two points by a path which intersects {\bf only} (at most) the submanifold of  ``simply critical'' admissible simple Boutroux curves ${\mathbf \Sigma}_{reg}^{0}$.

\subsection{Differential geometry of  $\mathfrak B_{reg}$: $\Pot$ as a local coordinate system}
We want to prove that $\Pot:\B\to \V$ is a local isomorphism in a neighborhood of a simple Boutroux curve, independently of any other condition (pre-admissibility, admissibility, criticality).

\bl[Local Isomorphism Lemma]
\label{lemmalift}
Let $[y_0]\in \B $ be a simple Boutroux curve with $\Pot[y_0] = (V_0,T_0)$.

Consider a neighborhood $\mathfrak P$ of $y_0^2 =P_0(x)$ consisting of Boutroux curves  $ y^2 =  P(x)$ whose zeroes are in a suitably small  polydisk around the zeroes of $P_0(x)$ and of the {\em same} multiplicity (and also the leading coefficient of $P$ is in a neighborhood of that of $P_0(x)$).
 
Then  $\Pot: \mathfrak P \to \V$ is a local isomorphism. 
In particular each cell (Def. \ref{cell})  of  $\B_{reg}$  ($\B_{reg}^{<adm}$,   $\B_{reg}^{adm}$ ) is a manifold of real dimension $2d+3$ and $\Pot$ gives local coordinates to it, inducing a $\mathcal C^\infty$ manifold structure that coincides with that induced by the coefficients of  $\mathbf P$.
\el
{\bf Proof.}
The last statement follows from the first and the following observation: if a curve $[y]\in \B $ is simple and noncritical (hence $[y]\in \B_{reg}\setminus \mathbf \Delta$) then all the  Boutroux curves  $[\wt y]$ in a neighborhood of $[y]$ consisting of simple curves  have zeroes of the same multiplicity. Indeed double zeroes cannot be split into two simple zeroes because then the newly created branchpoints would necessarily not belong to the zero levelset $\mathfrak X$ (a simple continuity argument on $h(m)$) and hence the curve immediately would fail the Boutroux condition.


Consider the neighborhood of $P_0(x)= C_0 \prod_{j=1}^\mu (x-m_{j,0})^2 \prod_{j=1}^{2g+2} (x-\a_{j,0})$ consisting of polynomials of the same form and with the zeroes nearby the ones of $P_0$.

The Boutroux conditions are $2g+1$ transcendental constraints on the positions of the zeroes of $P(x)$ and the leading coefficient;
\be
\mathcal F_\gamma (P) := \Re \le(\oint_{\gamma} y\d x \ri) = 0, \  \ \gamma \in H_1(\L\setminus \{\infty_\pm\}).
\ee
Viewing the $\mathcal F_\gamma$ as functions of the zeroes of $P$ it is clear that they are real-analytic. The isomorphism will follow from the implicit function theorem if we prove that the Jacobian is invertible.

To do this it is sufficient to prove that the push-forward by $\Pot$ of the tangent space to the constraints $\{\mathcal F_\gamma\}$ within the class of polynomials $P(x)$ with zeroes of constant multiplicity is  surjective and injective in the tangent space of $\V$.

Thus suppose $P_t $ is a one-parametric  deformation of $P$  (simple curve) preserving the Boutroux condition {\em and} the multiplicities of the double zeroes; thus $\dot P = \frac \d{\d t} P_t$ is a polynomial of the same degree $2d$. 
We write the deformation family as 
\be
P_t(x)  = C_t\prod_{k=1}^{\mu} (x-m_k(t))^2 \overbrace{\prod_{j=1}^{2g+2}(x-\a_j(t))}^{:={w_t}^2}=  {M_t }^2(x) w_t^2\ .
\ee
where the roots depend on $t$.
The potential/charge undergo a variation $\delta (V,T) = \delta t(\dot V,\dot T)$ given by 
\be
\dot V(x) = \res{\infty} \frac {\dot P(x)\d x }{y}\ ,\ \ \ \dot T = \res{\infty} \frac {\dot P(x)\d x}y
\ee
while $\dot P(x)$ must satisfy the {\em infinitesimal Boutroux condition}
\be
\oint_\gamma \frac {\dot P(x)\d x}{y} \in i\R\ ,\ \ \ \forall \gamma \in H_1(\L\setminus \{\infty_\pm\})\ ,\label{infinitesimal}
\ee
Note that necessarily $\dot P_t $ is divisible by $M_t$ (because the double roots remain such along the motion) and hence we can write it as follows
\be
\dot P_t (x)  = M_t (x)  R(x) \ \ \Rightarrow \ \ \oint_{\gamma} \frac { R(x)} { w_t }\d x\in i\R
\ee
with the degree of $ R(x)$ at most $g+d+1$. In intrinsic terms the differential 
\be
\dot \omega:= \frac { R(x)} { w_t}\d x
\ee
is a meromorphic differential on the hyperelliptic Riemann surface of $w_t$ with purely imaginary periods and poles only above $x=\infty$. Standard theory \cite{FarkasKra,fay} implies that it is uniquely determined by its singular part at the poles. In other words
\be
\dot V_t'(x)  -\frac { \dot T_t}x + \mathcal O(x^{-2}) = \frac { R(x)}{w_t}
\ee
uniquely determines the differential $\dot \omega$ under the requirement that all periods are imaginary. This means that the tangent space to the space of simple Boutroux curves is naturally isomorphic to the tangent space of external potentials and total charges, of real dimension $2d+3$ (the $d+1$ complex coefficients of $V'$ and the real total charge $T$).
\footnote{Note that the motion of the roots of $P_t(x)$ are determined by  
\be
\dot m_k(t) = -\frac {\dot P'(m_k(t))}{P_t''(m_k(t))}\ ,\qquad
\dot \a_j(t) = -\frac {\dot P(\a_j(t))}{P_t'(\a_j(t))}\ .\label{moveroots}
\ee}
{\bf Q.E.D.}\par \vskip 5pt

In the course of the proof we have shown that the tangent space $T_{(V,T)}\V$ for $(V,T)\not\in \mathbf \Sigma$  is naturally isomorphic to the space of second--kind differentials on $\L$ of the form 
\be
\dot \omega = \frac{ R(x)\d x}{w} \ \ \deg R(x) \leq g+d+1\ , 
\ee
that have poles  of order $d+2$ above $x=\infty$ and normalized to have purely imaginary periods. This space and this identification deserve a 
\bd
\label{Identifica}
Let $P(x)$ be a polynomial of degree $2d$ defining a (nodal) hyperelliptic curve of the form 
\bea
y^2 = P(x)&\&  := C^2 \prod_{j=1}^\mu (x-m_\ell)^{2k_j} \prod_{j=1}^{2g+2} (x-\a_j)^{2r_j+1}=  M(x)^2 \ w^2\\
 &\& w^2 := \prod_{j=1}^{2g+2} (x-\a_j)\ , \ \ M(x) :=  C^2 \prod_{j=1}^\mu (x-m_\ell)^{2k_j} \prod_{j=1}^{2g+2} (x-\a_j)^{2r_j}\ ,\ \ k_j, r_j \in \N\ .
\eea
Let  $\H_{\Im}(y)$ denoted the linear space of meromorphic differentials over the Riemann surface of $w$ 
\begin{itemize}
\item
with poles of order not greater than $d+2$ above $x=\infty$;
\item  skew-symmetric w.r.t. the hyperelliptic involution;
\item with zeroes at least of order $k_j-1$ at $m_j$ and $2r_j$ at the branchpoints $\a_j$;
\item with purely imaginary periods,
\end{itemize}
hence of the form 
\bea
\dot \omega :=\prod_{j=1}^\mu (x-m_\ell)^{k_j-1} \prod_{j=1}^{2g+2} (x-\a_j)^{r_j} \frac {S(x) \d x}{w}\ ,\ \ \ 
\deg S = g+1+d-\delta \ ,\cr
\ \delta:= \sum_{j=1}^\mu (k_j-1) + \sum_{j=1}^{2g+2} r_j\ ,\ \ d = g+1 +\mu +\delta  
\eea

We define the map
\be
\begin{array}{ccl}
  \Pot_\star :\H_{\Im}(P)&\longrightarrow& T_{\Pot[y]}\V \\
 \dot \omega & \longrightarrow &\Pot_\star(\dot \omega)  = (\dot V(x), \dot T)
\end{array} 
\ee
where $\dot V(x)$ is a constant--free polynomial and $\dot T$ a (real) number  identified by 
\be
\frac {S(x) \d x}{w} \sim \le(\dot V'(x) - \frac {\dot T} x + \mathcal O(x^{-2})\ri)\d x\ .
\ee
\ed
We  have already shown that if $\delta =0$ (i.e. the curve is simple, $k_j=1, r_j=0$) this map is an isomorphism between $\H_{\Im}(P)$ and $T_{\Pot[y]} \V$ (in fact, irrespectively of $P(x)$ satisfying the Boutroux condition or not). 

If $\delta \geq 1$ then this map is still injective (as follows from standard theory \cite{FarkasKra}) but not surjective and the  (real) rank is $2d+3-2\delta$; thus $\delta$ is a ``defect'' or {\em codimension}. We will use $\Pot$ and $\Pot_\star$ only for simple (Boutroux) curves, however, so the definition is more general than strictly needed.

\subsection{Differential geometric description of ${\mathbf \Sigma}_{reg}^0$ and transversality}
\label{discriminantgeometry}
Suppose a simple admissible Boutroux curve $[y_{cr}]\in \mathbf \Delta_{reg}^{0}\subset \B$ is  simply-critical and consider its image $(V_{cr}, T_{cr}) =\Pot[y_{cr}] \in{\mathbf \Sigma}_{reg}^0\subset  \V$. Recall that this means that a single double zero $m$ of $y^2 = P(x)$ lies on the zero levelset $\mathfrak X$. 

We are going to show below (Prop. \ref{smoothDisc}) that $\mathbf \Delta_{reg}^0$ is also locally a manifold and $\Pot$ gives a (local) isomorphism with its image (a smooth hypersurface in $\V$).

Note that a simply critical Boutroux curve $[y_{cr}]\in \mathbf \Delta_{reg}^0$ lies on the boundary of two or more cells (Def. \ref{cell}) of $\B_{reg}$, which (recalling Lemma \ref{lemmalift}) are all smooth manifolds of the same (real) dimension $2d+3$.

 We will need this in order to formulate the notion of {\bf transversality} to the discriminant. More precisely
\bp
\label{smoothDisc}
Let  $[y_{cr}] \in {\mathbf \Delta}_{reg}^0$  be simple, admissible and simply-critical. Then there exists a neighborhood $U\subset \B^{adm}$ of $[y_{cr}]$ such that  $\Pot \le( U \cap {\mathbf \Delta}_{reg}^0\ri) \subseteq {\mathbf \Sigma}_{reg}^0$ is a smoothly embedded surface of real codimension $1$.
\ep
{\bf Proof.}
Let $m_{cr}$ be the critical double zero of $y_{cr}^2 = P_{cr}(x)$. If $y^2  = P(x)$ is in a small neighborhood of $P_{cr}(x)$, the constraints that imply that $[y]\in {\mathbf\Delta}_{reg}^0$ are that a double root $m$ (near $m_{cr}$) still belongs to $\mathfrak X$ and hence 
\be
\int_{\a}^m y\d x \in i\R\ ,\ \ \oint_{\gamma} y\d x \in i\R\ ,
\ee
where $\a$ is any branchpoint.
Again, these are $(2g+1)+1$ real smooth constraints on the coefficients of $P(x)$, hence the codimension count is immediate.

The tangent space to these constraints --similar to the proof of Lemma \ref{lemmalift}-- 
is the subspace of $\H_\Im(P)$ (Def. \ref{Identifica})  cut by the additional linear equation
\be
\mathcal F[\dot \omega] = \int_\a^m \dot \omega  \in i\R\ .
\ee 
We only need to prove that this constraint is linearly independent of the infinitesimal Boutroux conditions (\ref{infinitesimal}) (i.e. the vanishing of the real part of all  periods of $\dot \omega$).
It suffices to exhibit a differential in this space which has imaginary periods but nonzero integral from $\a$ to $m$. 
But this is obvious: for  the differential $ \d w$ is a differential which has all zero periods (it is exact!) and  pole of degree $g+2<d+g+1$, hence $\d w\in \H_\Im$. Moreover $\int_\a^m \d w = w(m)\neq 0$. Thus the differential we seek is --for example--
\be
\dot \omega = \frac {\d w}{w(m)}\ ,\ \ \int_{\a}^m \dot \omega = 1\ ,\ \ \oint_{\gamma} \dot \omega =0\ .
\ee
{\bf Q.E.D.}
\paragraph{Transversality}
This notion is implicit in the proof of Prop. \ref{smoothDisc} but it is worth making it explicit. A transversal path to ${\mathbf \Sigma}_{reg}^0\subset \V$ will be such that  the tangent vector $(\dot V, \dot  T)$ is  associated  to the second-kind differential with imaginary periods (via the identification of $T_{(V,T)}\V$ with the space of such Abelian differential of Def. \ref{Identifica}) with 
\be
\Re \int_\a^m\dot \omega = \rho \neq 0\ .\label{transversality}
\ee
\subsection{Self--intersections of ${\bf \Sigma}_{reg}^0$}
We point out that the proof of Prop. \ref{smoothDisc} is easily adapted to show that the self--intersection of ${\bf \Sigma}_{reg}^0$ are {\em completely transversal}.
Indeed if $[y_{cr}]$ is multiply simply critical (i.e. has several double roots $m_j$ on the zero levelset $\mathfrak X$), then the condition is described by the constraints 
\be
\Re \int_a^{m_j} y \d x=0, \ \ j=1\dots r.
\ee
These constraints are independent and hence transversal to each other: to show this it suffices to exhibit in infinitesimal form  a $\dot \omega \in \H_\Im(P)$ such that 
\be
\Re \int_\a^{m_j} \dot \omega_\ell = \delta_{j,\ell}\ ,\ \ \forall \ell = 1\dots r.
\ee
But this is simply achieved considering 
\be
\dot \omega _\ell = \frac {\d \le( \prod_{j\neq \ell} (x-m_j) w(x)\ri)}{\prod_{j\neq \ell} (m_\ell-m_j) w(m_\ell)}
\ee

For our purposes it will be immaterial to consider curves which are multiply simply critical, although in some circumstances (e.g. under some reality conditions for the potential) this may happen. However the analysis to follow would not seriously differ in this slightly more general case.
\br[The zero-charge locus $T=0$]
\label{zerocharge}
Suppose the total charge is zero $T=0$: since the total charge is the total mass of the measure supported on the branchcut structure, either this measure is signed and hence the curve is inadmissible or the curve is a perfect square of the form 
\be
y^2   = (V'(x))^2\ .
\ee
This locus is a multiple branching locus: in its neighborhood there are $d$ admissible curves of genus $0$ and ``small'' total charge, obtained by solving the algebraic equations for the coefficients of $P(x)$ given by 
\bea
y^2 = M^2(x) (x-\a_1)(x-\a_2)\ ,\ \ \ \deg M = d-1\\
M(x) \sqrt{(x-\a_1)(x-\a_2)} \sim V'(x) -\frac T x + \mathcal  O(x^{-2})\ .
\a_j = z_0 + \mathcal O(\sqrt{T})
\eea
where $z_0$ is a simple zero of $V'(x)$ chosen arbitrarily; here $V'(x)$ is assumed to have simple zeroes for otherwise we have a particularly singular sublocus of $T=0$. The resulting Boutroux curve is certainly admissible for small $T$'s. 
\er

\subsection{Coalescence of simple roots: approaching the boundary of a cell }
Let $\B_{reg}^0$ be a cell of $\B_{reg}$ (Def. \ref{cell}); the considerations to follow extend {\em verbatim} if we restrict to (pre)admissible Boutroux curves.

A point $[y_{cr}]$ on the smooth boundary $\mathbf \Delta_{reg}^0$ has a single double root $m_{cr}$  on the zero levelset $\mathfrak X = h^{-1}(0)$; this can be the limit of two types of curves
\begin{enumerate}
\item  simple curves with a double zero $m$ that approaches $\mathfrak X$ or
\item  simple curves with two simple zeroes $m_\pm$ merging together.
\end{enumerate}

By the first part of Lemma \ref{lemmalift} the first case is in no way ``singular'': the map $\Pot$ gives a local coordinate system in a full neighborhood of $[y_{cr}]$.

Let us consider viceversa the second case: we will  show below that the differentiable structure induced on $\mathcal B_{reg}$ by $\Pot$ can be extended  to $\mathbf \Delta_{reg}^0\cap\overline{ \B_{reg}^0}$, extending thus the manifold $\B_{reg}^0$ to a manifold-with-boundary. Suppose we are given a vector field $(\dot V(x),\dot T)\in T\V$ (at least $\mathcal C^0$) defined in a neighborhood of $(V_{cr}, T_{cr})\in \mathbf \Sigma_{reg}^0$. Let
 \be
P_{cr}(x)  = (x-m_{cr})^2  M^2_{cr}(x) \prod_{j=1}^{2g+2} (x-\a_{j,cr})\label{critcurve}
\ee
be the corresponding critical Boutroux curve. 
 
 Consider simple noncritical Boutroux curves $[y]\in \B_{reg}^{0}$ of the form
\bea
y^2 = P(x) =  M^2(x)(x-m_+)(x-m_-)\prod_{j=1}^{2g+2} (x-\a_j) =: M^2(x) w^2(x)\cr
w^2 = (x-m_+)(x-m_-)\prod_{j=1}^{2g+2} (x-\a_j)\ .
\eea
 in a neighborhood $\mathfrak P$ of $P_{cr}$ such that  the coefficients of $M$ are in suitable small neighborhoods of those of $M_{cr}$ and all the roots are in small neighborhoods of homonymous critical ones; in particular $m_\pm$ are roots in a small neighborhood of $m_{cr}$.
Inside $\mathfrak P$ the discriminant is represented simply by the locus $\{m_+=m_-\}$ and contains our critical Boutroux curve as well.
The genus of $y^2 =P(x)$ is $g+1$ and we choose the Torelli marking $\{a_j, b_j\}$ so that $a_{g+1}$ is a small loop encircling both $m_\pm$ and $b_{g+1}$ the corresponding symplectic dual; this is a the same setting as in Chapter 3 of \cite{fay}.

We need to show that $\dot \omega = \frac {S(x)}w = \Pot_\star^{-1}(\dot V(x),\dot T)$ (Def. \ref{Identifica}) is continuous as we approach the boundary $m_+=m_-$ (the inverse is well defined since $\Pot_\star$ is an isomorphism for simple curves).
 Note immediately that $S(x)$ depends linearly on $(\dot V(x),\dot T)$  and hence has the same smoothness class.

We need to show that 
\begin{enumerate}
\item the coefficients of $S(x)$ remain bounded as $[y]\to [y_{cr}]$;
\item the limiting vector field $\dot \omega_{cr}  = S_{cr}/w_{cr}\d x$ is  {\em holomorphic} also at $m_{cr}$, namely that $S_{cr}$ has a simple zero at $m_{cr}$. In this way the tangent space will coincide with the tangent space described in Prop. \ref{smoothDisc} for the tangent to the discriminant, allowing us to conclude that the differentiable structure extends to the boundary.
\end{enumerate}
We construct $S(x)$  as follows. Start from 
\be
S_0(x) = \le[w(x) \le(\dot V'(x)  - \frac {\dot T}x\ri)\ri]_+ - C\label{ssss}
\ee
where the subscript denotes the polynomial part;
here $C$ is a constant such that $S_0(m_+)\equiv 0$ if the genus of $w$ is greater than one, or else it is zero\footnote{Strictly speaking the introduction of this constant would be unnecessary, but it simplifies some considerations to follow.}.
 This is a polynomial expression in the roots of $w$ and hence regular as long as none of the roots of $w$ goes to infinity. By assumption, since $(\dot V,\dot T)$ is at least $\mathcal C^0$, then $S_0(x)$ is at least $\mathcal C^0$.

We have now a meromorphic differential 
$
\eta := \frac {S_0(x)\d x} w
$
which has the prescribed polar expansion\footnote{Note that any constant $C$ (or addition of any polynomial of degree $\leq g-1$ for that matters) in (\ref{ssss}) would not change the singular behavior of $\eta$ at infinity if $g\geq 1$.} but  whose  periods are not imaginary, except if the genus of $w$ is zero, in which case there is nothing else to do. If the genus is higher than zero, we must add a linear combination of the holomorphic differentials in order to achieve this. The first goal is to show that the resulting vector field $\dot \omega =\frac { S(x)} w \d x $ will have bounded and continuous coefficients (namely that the coefficients of $H_P$ are bounded and continuous).

Let $\omega_j,\ j=1,\dots g+1$ be the normalized Abelian differentials of the first kind \cite{FarkasKra} and $\tau_{jk} = \mathbb A_{jk} + i \mathbb B_{ji}$ be the corresponding period matrix. Recall that $\mathbb B_{jk}$ is {\em positive definite}.

Suppose $m_+\to m_-$; it is proved in \cite{fay}  that $\mathbb A$ is $\mathcal O(1)$ and all coefficients of $\mathbb B$ are also bounded except 
\be
\mathbb B_{g+1,g+1} = \frac 1{4\pi} \ln |m_+-m_-| + \mathcal O(1).
\ee
and the determinant of $\mathbb B$ behaves thus
\be
\det[ \mathbb B]_{j,k\leq g+1} = \frac 1{4\pi} \ln |m_+-m_-|  \det[\mathbb B]_{j,k\leq g} + \mathcal O(1)
\ee
where the subdeterminant tends to a {\em nonzero} constant; indeed \cite {fay} the period matrix $\tau$ has the expansion
\be
\tau = \le[ \begin{array}{c|c}
\tau_{jk} +\mathcal O(\delta) & \vec A + \mathcal O(\delta)\\[10pt]
\hline
\ds \vec A^t + \mathcal O(\delta) & \ds \frac 1 {4i\pi} \ln \delta  + \mathcal O(1)
\end{array}\ri]\ ,\ \ \ \ \delta := m_+-m_-
\label{periodmatrix}
\ee
where the constants $\vec A$ can be expressed in terms of the first kind differentials (see \cite{fay} for even more detailed information) and $\tau_{jk}$ in the principal $g\times g$ block is the period matrix for  the limiting curve $\mu^2 = \prod_{j=1}^{2g} (x-\a_j)$.
This holds independently of any Boutroux condition\footnote{In comparing with \cite{fay}, pag. 50 and following, the reader should keep in mind that Fay uses a different normalization for the period matrix, in which the real and imaginary parts are interchanged.}.
If we denote by $\mathcal A_j =\Re  \oint_{a_j} \eta$ and $\mathcal B_j = \Re \oint_{b_j} \eta$ the periods of $\eta$,  the normalized differential is thus 
\be
\eta - \vec \omega^t \le[ \vec{\mathcal A}  - i \mathbb B^{-1}\le( \mathbb A \vec {\mathcal A} - \vec{\mathcal B}\ri)\ri] = \frac{ S(x)\d x}{w}\ .
\ee
 Now all the periods of $\eta$ are bounded: also the  pinched cycle ($b_{g+1}$)  (which we can think of as a loop encircling the root $m_+$ and some other root but $m_-$)
 is bounded because of our choice of the constant $C$ in (\ref{ssss}).

The last column of $\mathbb B^{-1}$ is  $\mathcal O\le (\frac 1{\ln| \delta|}\ri)$ in all the entries except for ${\mathbb B^{-1}}_{g+1,g+1}$ which is just $\mathcal O(1)$.
Note that the ambiguity in the choice of symplectic basis --which is ultimately responsible for the multivaluedness of the expansion of the period matrix by Lefschetz theorem on vanishing cycles-- does not affect the definition of the polynomial $S$ because the infinitesimal Boutroux condition is independent of the choice of basis in the homology of the curve.\footnote{It is also shown in \cite {fay} that $\omega_{g+1}$ tends to the normalized third kind--differential on the de-singularization of the limiting hyperelliptic curve.} This concludes the proof of the continuity of $\dot \omega (x)$ in $\mathfrak P$.
We still need to prove, however, that 

\bl
The limiting $\dot \omega_{cr}$ has a zero at $m_{cr}$.
\el
{\bf Proof.}
Considering the vanishing cycle  $\gamma_v$  (a loop encircling $m_\pm$) we see that the Boutroux condition implies 
\be
\oint_{\gamma_v} \dot \omega_{cr}  = \oint_{\gamma_v} \frac{S_{cr}(x)} {(x-m_{cr}) \sqrt{\prod_{j=1}^{2g+2} (x-\a_{j,cr})}}  =  2i\pi \frac {S_{{cr}}(m_{cr})}{\sqrt{   \prod_{j=1}^{2g+2} (m_{cr}-\a_{j,cr}) }}\in i\R
\ee
On the other hand, since this is the limit of Boutroux curves, considering the Boutroux condition on the pinched cycle (a loop encircling one of $m_\pm$ and another simple root), we get 
\be
\oint_{\gamma_p}  \dot \omega_{P}(x)   \sim   \frac {S_{{cr}}(m_{cr})}{{\sqrt{   \prod_{j=1}^{2g+2} (m_{cr}-\a_{j,cr}) }} } \ln |m_+-m_{cr}|
 + \mathcal O(1)\in i\R
\ee
Thus 
\be
 \frac {S_{{cr}}(m_{cr})}{{\sqrt{   \prod_{j=1}^{2g+2} (m_{cr}-\a_{j,cr}) }} }
 = \{0\}\ . \ \ \ \ {\rm \bf Q.E.D.}
\ee
Since we now know that $S_{cr}(m_{cr})=0$ and given the continuity proven above, we also know that the limit of the value of a pinched cycle is finite. Note that the polynomial $S(x)$ has an expansion of the type
\be
S(x) = S_{cr}(x) + \epsilon S_1(x) + \dots
\ee
where $\epsilon$ is some infinitesimal of $|m_+-m_-|$ which we are set to estimate; in principle this is encoded in the expansion of the period matrix  (\ref{periodmatrix}) but it is simpler to reason as follows:
the value of the pinching cycle is 
\bea
\frac 1 2\oint_{\gamma_p} \frac {S(x)\d x}w &\& = \int_\a^{m_+}  \frac {S(x)\d x}w = \int_{\a_{cr}}^{m_{cr}} \frac {S_{cr}(x)\d x}{w} + \epsilon   \int_{\a_{cr}}^{m_{+}} \frac {S_{1}(x)\d x}{w} + \dots = \cr
&\& =  \int_{\a_{cr}}^{m_{cr}} \frac {S_{cr}(x)\d x}{w} + \epsilon   \ln|m_+-m_{-}|  \frac {S_{1}(m_{cr})}{
\sqrt{   \prod_{j=1}^{2g+2} (m_{cr}-\a_{j,cr}) }} + \dots
\eea
Taking the real part of the above and setting it to zero we conclude that we must have 
\bea
\epsilon = \frac 1{\ln|m_+-m_-|}\ ,\ \ 
\Re \le(\int_{\a_{cr}}^{m_{cr}} \frac {S_{cr}(x)\d x}{w_{cr}}\ri) = \Re \le(  \frac {S_{1}(m_{cr})}{
\sqrt{   \prod_{j=1}^{2g+2} (m_{cr}-\a_{j,cr}) }} \ri)
\label{bingo}
\eea
Note that the LHS of (\ref{bingo}) is a well-defined integral because the $S_{cr}(x) = (x-m_{cr})R(x)$ and $w_{cr} = (x-m_{cr}) w^0$, so that the pole in the denominator is canceled by the numerator. 

On the other hand, repeating the computation of the vanishing period now that we know that $S_{cr}(m_{cr})=0$ and considering the next order we find 
\be
\oint_{\gamma_p}  \dot \omega_{P}(x)   \sim \frac {2i\pi}{ |\ln |m_+-m_-||} \frac {S_{1}(m_{cr})}{
\sqrt{   \prod_{j=1}^{2g+2} (m_{cr}-\a_{j,cr}) }}\in i\R
\ee

We see then that the expression $  \frac {S_{1}(m_{cr})}{
\sqrt{   \prod_{j=1}^{2g+2} (m_{cr}-\a_{j,cr}) }}$ must be {\em real} and must match the LHS of eq. (\ref{bingo}).  
We have thus proved 
\bp
\label{propbingo}
The continuation of a tangent vector $\dot \omega$ near a critical Boutroux curve obtained by coalescing two roots $m_+, m_-$ into a critical double root $m_{cr}$ has the expansion 
\bea
\dot \omega &\&= \dot \omega_{cr}  + \frac 1{\ln |m_+-m_-|} \le(\dot \omega_1+ o(1)\ri)\cr
&\&  \dot \omega_{cr}  = \frac {R(x)}{\sqrt{\prod_{j=1}^{2g+2} (x-\a_{j,cr})}},\ \ 
\dot \omega_1 (x) = \frac{ S_1(x)}{(x-m_{cr}) \sqrt{\prod_{j=1}^{2g+2} (x-\a_{j,cr})}} 
\eea
where $\dot \omega_{cr}$ is holomorphic at $m_{cr}$  and 
\be
 \R \ni  \frac {S_{1}(m_{cr})}{
\sqrt{   \prod_{j=1}^{2g+2} (m_{cr}-\a_{j,cr}) }} = \Re\le(\int_\a^{m_{cr}}  \dot \omega_{cr} \ri).
\ee
\ep
\section{Admissible Boutroux curves and connectivity}
\label{secfive}
\subsection{Precise statement of the problem}
We now come back to the setting described in the introduction: we are given an {\bf external potential} $V(x)$ (complex polynomial) and a {\bf total charge} $T>0$ together with the combinatorial data described below, encoding the information of the contours $\gamma_\ell$ (see Fig. \ref{contours}) actually used in the definition of the orthogonal polynomials.

\begin{figure}
\resizebox{13cm}{!}{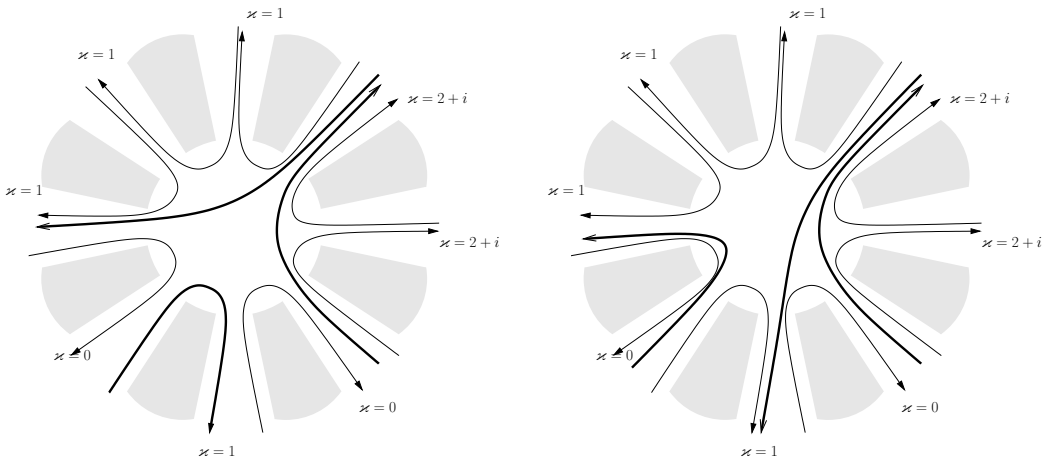}
\caption{Using the same example of Fig. \ref{contours} with some specific values of the currents $\varkappa_\ell$ (indicated near the tip of the arrows). The sectors with zero total incoming current can be skipped by the irreducible connectivity pattern and the contours carrying zero current are irrelevant. In this example there are two notably distinct irreducible connectivity patterns because the highways carry the same total traffic.}
\label{totalcurrent}
\end{figure}

Using the notion of total incoming current (see caption of Fig. \ref{contours}) we can determine which sectors are actually ``used'' in the definition of our moment functional $\mathcal L$: indeed if the total current for sector $\mathcal S_{\ell_0}$ is zero, this means that we can use Cauchy theorem to eliminate the two half-branches of the two contours $\gamma_{\ell_0}, \gamma_{\ell_0+1}$ (indices are modulo $d+1$) accessing it combining into a contour that ``skips'' the sector (see Fig. \ref{totalcurrent}).

\br
The fact that the currents form an additive group is also the reflection of the fact that the Riemann--Hilbert problem for the generalized orthogonal polynomials has only upper triangular Stokes' matrices, which form an Abelian group.
\er

We proceed this way until only sectors with nonzero total incoming/outgoing current are left. The data of the surviving contours together with  the traffics (or currents) carried by them  and which sectors they connect will be referred to as the (irreducible) {\bf connectivity pattern} (denoted by $\wp$). 

\br[Non-uniqueness of the irreducible connectivity pattern]
The connectivity pattern is in general not unique: a different rewiring of the sectors must be such that 
\begin{itemize}
\item If there are $K+1$ sectors with nonzero net current then we use the minimal number of  {\bf wires} (or {\bf highways}), that is, $K$.
\end{itemize}
If, in particular, some highways carry the same traffic, then the highway system can be rerouted in several ways. An example is in Fig. \ref{totalcurrent}.
\er

\bd
\label{defconn}
For a given admissible Boutroux curve, two sectors $\mathcal S_j, \mathcal S_k$ are {\bf connectible} if there exists a   continuous piecewise smooth path $\gamma:\R\to \C$  (called the {\bf connecting path}) such that one end belongs asymptotically to sector $\mathcal S_j$ and the other to sector $\mathcal S_k$ and such that 
\be
h(x)\geq 0\ ,\ \ x\in \gamma.
\ee
\ed

The reader should visualize that each emerged continent borders at infinity one or more sectors $\mathcal S_j$ and continents (and thus sectors) may (or may not) be joined by a path that never touches the water but can use the causeways: such a path may contain some arcs from the branchcut structure $\mathcal B$ or lie entirely above water (in which case it is an {\bf elevated highway}).
\bd
In an irreducible connectivity pattern, given two highways with the same traffic connecting two disjoint pairs of sectors we call a {\bf exchange re-routing} (Fig. \ref{totalcurrent}) the re-routing of the two highways  in such a way that they connect the four sectors in the other way compatible with the total flow of traffic from the sectors.
\ed
\bd
\label{defcompatible}
Let it be given  an irreducible connectivity pattern $\wp$ (with specified traffics $\varkappa$) and an admissible Boutroux curve with branchcut structure (causeways) $\mathcal B$.  We say that the connectivity of the Boutroux curve is {\bf compatible} with the given pattern $\wp$ (or briefly that the Boutroux curve is compatible with $\wp$) if 
\begin{itemize}
\item each highway connecting two sectors can be smoothly deformed to a connecting path (as in Def. \ref{defconn}) up  to an exchange re-routing with another highway of the same traffic;
\item  each causeway after  the re-routings carries some {\em nonzero} net traffic.
\end{itemize}
\ed 
The notion of compatibility in Def. \ref{defcompatible} does not exclude that the Boutroux curve in question allows a connecting path (Def. \ref{defconn}) between two sectors {\em without} incoming traffic in the irreducible pattern $\wp$: this can happen as long as the connecting path is  {\em strictly elevated} (i.e. does not visit any causeway) and connects the two sectors. In other words a Boutroux curve may be compatible with more than one irreducible connectivity pattern.

After this lengthy sequence of definitions and preparations we finally state precisely the problem we want to solve.
\bt[Main theorem]
\label{maintheorem}
 For any   $(V(x),T)\in \V \setminus \mathbf \Sigma_{sing}$ and any connectivity pattern $\wp$  there exists a unique simple (possibly critical) admissible Boutroux curve compatible with the  given connectivity pattern $\wp$\footnote{If the potential is non generically chosen then the curve may fail to be simple and some degeneration may occur. Of course the existence is not seriously affected since one may use a continuity argument.}.
\et
The proof of this theorem will be embodied by the rest of the paper: the main idea is that of using a deformation argument starting from a ``base'' Boutroux curve with the desired connectivity pattern but not necessarily with the same external potential. We then prove that we can smoothly deform it so as to reach the desired external field $V(z)$ without decreasing the connectivity pattern.

There are two main ingredients to this program: 
\begin{enumerate}
\item showing that for each connectivity pattern there is an admissible Boutroux curve with that pattern (Section \ref{secbasecurve});
\item showing that we can deform such curve until the external potential matches our chosen one without severing any of the highways that carry net traffic between sectors. The highways follow continuously the deformation (while connecting the same two sectors) up to an exchange re-routing (if allowed by the traffics). This is contained in Sections \ref{secdeform}, \ref{sectsimple}, \ref{sectsing}, \ref{secunique}
\end{enumerate}

\subsubsection{Construction of a simple admissible Boutroux curve of given connectivity pattern}
\label{secbasecurve}
We take an irreducible form of the desired connectivity pattern and regard it as as a diagram. By construction the highways connect distinct sectors: an admissible sector $\mathcal S_j$ must necessarily be contained in an emerged continent since $h(x)$ behaves as $\Re(V(x))\to + \infty$ near $x=\infty\in \C P^1$. 

\begin{figure}
\resizebox{12cm}{!}{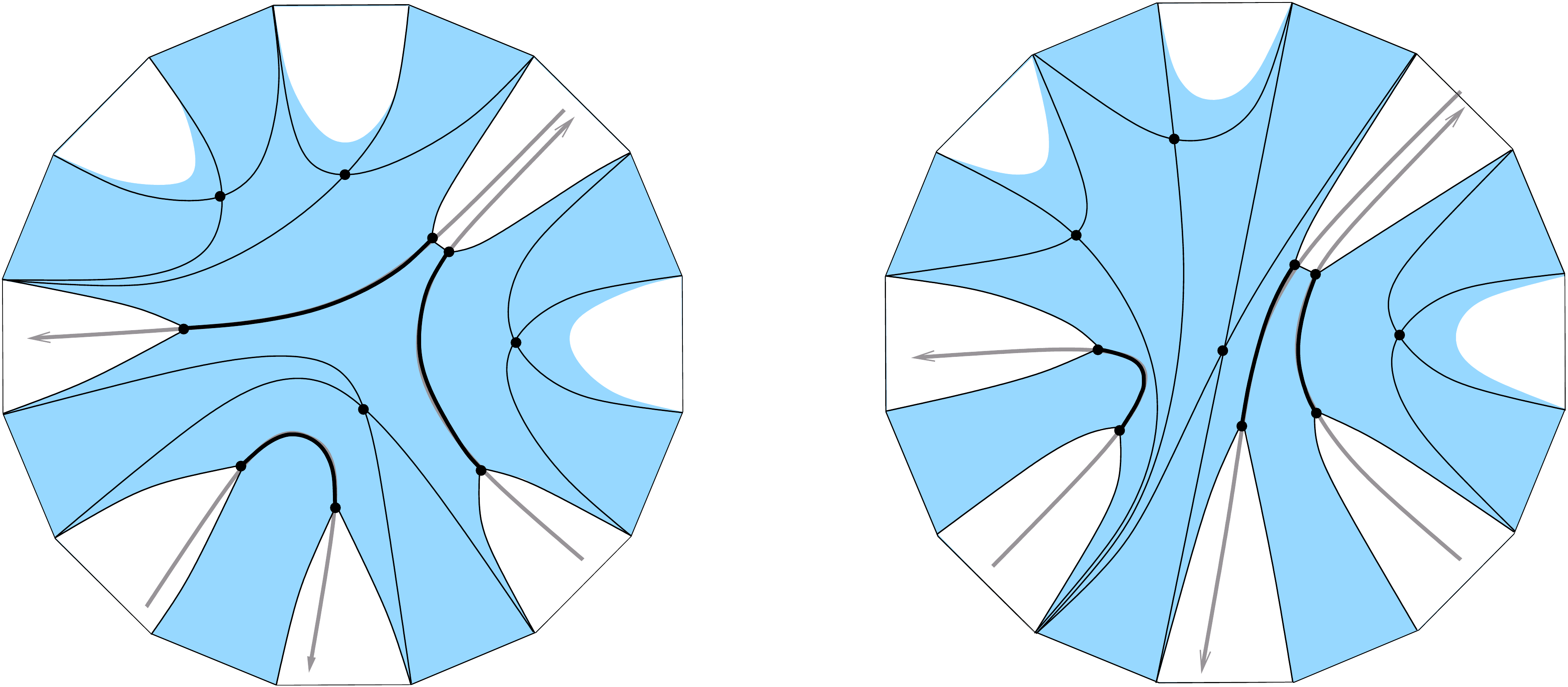}
\caption{The construction of the causeways/oceans from the irreducible connectivity pattern. The two clock diagrams correspond to the two equivalent irreducible patterns in Fig. \ref{totalcurrent}. Since two highways have the same traffic (see Fig. \ref{totalcurrent}) we have the two possibilities above. The two cases correspond to an {\em exchange re-routing}.}
\label{flooding}
\end{figure}

The steps to construct the clock-diagram of the admissible Boutroux curve are now
\begin{enumerate}
\item
Choose a highway and two points (entry/exit) on it,  and erase the two outer legs of the highway: the remaining part will be carried by a causeway though an ocean  and the entry/exit points will be simple zeroes. 
\item
From each entry/exit point we originate two shorelines that go to infinity in the clock-diagram along the two consecutive critical directions that limit the sector that the original highway was accessing.
\item
If a formed shoreline intersects the course of a different highway accessing the same sector, then we put the entry (or exit) of that other highway at the intersection point.
 We proceed to exhaustion until no more complete highways have left. If there were $K \leq d$ highways originally, there are now $2K\leq 2d$ entry/exit points. We thus have to arrange  $(2d-2K)/2 = d-K$ simple saddle points in order to have an admissible clock diagram.
 \item For each sector without accessing highway we draw a smooth (noncritical) shoreline bounding a unique continent containing that sector. We flood all the rest with water.
\item 
The shorelines/causeways constructed so far limit a certain number of oceans and each ocean has a finite number $M$  of connected boundaries (some of which may include a causeway): from the discussion in Sect. \ref{sectConstr} and point (\ref{morse}) we need to place $M-1$ saddle points below the sea-level and arrange suitably the four critical level lines from them. We leave to the reader to verify that this can be done respecting the rules of a clock-diagram. An example of the process is provided in  Fig. \ref{flooding}.
\end{enumerate}
Once the clock diagram has been constructed we can assign arbitrarily the decorations to the edges/strips and by the reconstruction theorem \cite{BertoMo} we know that this corresponds to an actual admissible Boutroux curve for some external potential $V_0(x)$ of degree $d+1$ and some total charge $T_0>0$. 
%
\subsection{Deformation of the curve towards the target potential}
\label{secdeform}
%
%
From Sect. \ref{secdim} we know that  $\B_{reg}^{adm}$ is coordinatized by $\Pot$. Moreover we claim the obvious 
\bp
Each cell of $\B_{reg}^{adm}$ contains Boutroux curves admissible with the same connectivity patterns. 
\ep
Indeed, by definition, a cell of $\B^{adm}$ consists of curves that can be deformed one into the other while in the process no saddle (double root) can cross a shoreline or causeway and no two simple roots can coalesce; hence the connectivity pattern cannot change.
\bt[Deformation theorem]
\label{mainthm}
Let $[y_0]\in \B_{reg}^{adm}$ (simple, noncritical) and  compatible with a given connectivity pattern $\wp$. 
Let $(V(x),T)\in \V\setminus \Sigma_{sing}$; then there exists an admissible Boutroux curve $[y]\in \Pot^{-1}(V,T)$ compatible with the same irreducible connectivity pattern $\wp$.
\et
The proof  is contained in Sections  \ref{sectsimple}, \ref{sectsing} along the idea expounded here below.
As outlined previously, the proof consists in fixing a smooth path $(V_t,T_t)$ from $\Pot[y_0]$ to $(V,T)$  (that avoids the co-dimension $2$ singular discriminant) and lifting this path into $\B_{reg}^{adm}$ in such a way that the connectivity pattern is preserved at all times: the crux of the proof is to show that, as we move from one cell of  $\B_{reg}^{adm}$ to another when crossing the discriminant we can {\bf choose} in which neighboring cell to lift the path in order to ensure the correct connectivity.
Namely the lift of $(V_t,T_t)$ must be such that 
\begin{itemize}
\item the algebraic curve remains Boutroux, simple and admissible at all times but possibly (simply) critical (Def. \ref{defsimple}) at some exceptional times ({\em phase transitions});
\item none of the highway is ever flooded or severed  in the process. Of course the issues are sudden changes in the topology of the connectivity (underlying the so--called  {\em nonlinear Stokes' phenomenon} \cite{ItsKapaev, ItsBook}), such as when a saddle emerges from, or sinks into, the ocean as a result of these ``tectonic shifts''.
\end{itemize}
%

%

Since by assumption the path $(V_t,T_t)$ may only cross $\mathbf \Sigma_{reg}^0$  the lift may cross $\mathbf \Delta_{reg}^0$, the set of simply--critical admissible Boutroux curves.
There are only two ways, as discussed in Sect. \ref{secdim} in which we can reach the boundary of a cell of $\B_{reg}^{adm}$ in the regular discriminant:

\begin{enumerate}
\item \label{unoa} if two branchpoints (connected by a critical line) merge into a double root; 
\item\label{due}  if a double root $m_t$ intersects the zero-level set $\mathfrak X$, namely if $h(m_t)=0$ at some time $t$.
\end{enumerate}
We will call these events {\em simple phase transitions}.
\subsection{Simple phase transitions}
\label{sectsimple}

 A simply critical, simple Boutroux curve $y^2 = P(x)$  must have a single double zero $m$ on the levelset $h^{-1}(0) = \mathfrak X$; it may then either belong to one or more causeways or  belong to a shoreline.

 \begin{wrapfigure}{r}{0.35\textwidth}
 \begin{center}
 \vspace{0cm}
 \includemovie[
   autoplay, autostop,repeat,toolbar,rate=2
 ]{3cm}{3cm}{Genus1.avi}
 \includemovie[
    autoplay, autostop,repeat,toolbar,rate=2
 ]{3cm}{3cm}{Genus1-0.avi}
 \end{center}
 \caption{These animations show two lifts of the same path in the $\V$ space. The potential is $V(x) = \frac {x^3}3 +  x$ and $T$ goes from $T=3$ to $T=1$. The simple transition is at $T=2$.}
 \end{wrapfigure}

The double root $m$  may belong to \\
{\bf (c0)} a  shoreline (i.e. all critical trajectories issuing from $m$ are shorelines);\\
{\bf (c2)} two causeways necessarily forming an angle of $\pi/2$ (i.e. two critical trajectories are shorelines and two are causeways);\\
{\bf (c4)} four causeways.

Each situation corresponds to points $[y]\in \B^{adm}$ at the boundary between several cells of  $\B^{adm}$, and the main point is to show that we can always choose to lift the path in a suitable cell with the appropriate connectivity pattern. 

In fact we will show that 
\begin{enumerate}
\item Case {\bf (c0)} is on the boundary of {\bf three} cells of  $\B_{reg}^{adm}$;
\item Case {\bf (c1)} is on the boundary of {\bf three} cells of $\B_{reg}^{adm}$;
\item Case {\bf (c2)} is on the boundary of {\bf four} cells of $\B_{reg}^{adm}$.
\end{enumerate}

\br
As a side note, each simply critical Boutroux curve is on the boundary of four cells of $\B$, but in general not all of them consist of admissible Boutroux curves. In particular in cases {\bf (c0,c1)} there is a fourth cell nearby, but it consists of inadmissible curves.
\er

In Section \ref{secdim} we showed that we can add the simply critical Boutroux curves  on the boundary of each cell of $\B_{reg}^{adm}$ turning them into (open) manifolds-with-boundaries. The whole point of the discussion to follow is to determine which two such components incident on the same parts of $\mathbf \Delta$ can be glued together to make a manifold without boundary. The issue is that the a ``left'' part can be only joined to a ``right'' part, and this allows to continue the lift of a path $(V_t,T_t)$ crossing the discriminant.

We are going to examine each case in the following paragraphs: each paragraph is accompanied by a marginal picture showing  the local connectivity of the neighboring cells.

 The dotted lines  show  which cell can be glued to which other and hence how the lift of a path lying on a cell can be lifted across the discriminant. Some paragraphs are also accompanied by short animations showing numerical examples\footnote{They can be viewed using Acrobat Reader version 7 or higher for  Windows or Mac. The Linux version of the reader does not allow inline viewing, but it should be possible to extract the animations and play them in an external viewer.}.

From {\em some}  situations we have two possible evolutions preserving admissibility through criticality depending on the decision of splitting or not the double root; which evolution one chooses depends on which connectivity pattern must be preserved. 

\paragraph{ Case {\bf (c0)}}
This can be the result of a sinking/emersion or a merging; if we arrived at {\bf (c0)} from an emersion resulting from the lift of $(V_t,T_t)$ up to $t=t_0$ then, in order to maintain the differentiability of the path $(V_t,T_t)$ we need to either continue the emersion of the saddle (i.e. do no split it) or split it into two branchpoints.  
\begin{wrapfigure}{r}{0.35\textwidth}
\vspace{-0.8cm}
\begin{center}
\resizebox{4.5cm}{!}{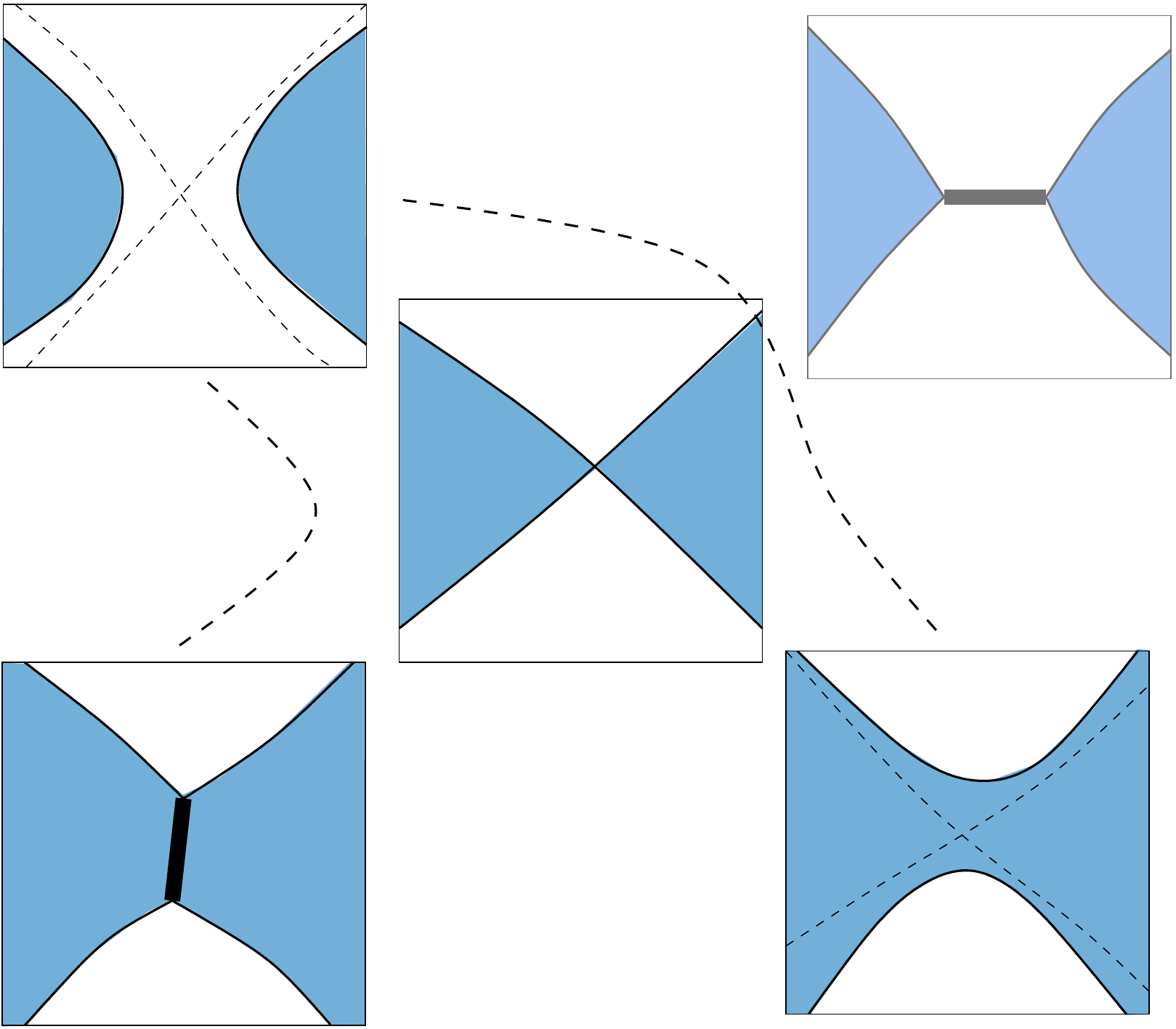}
\end{center}
\vspace{-1cm}
\end{wrapfigure}
If we let the saddle emerge ($2\mapsto 0\mapsto 3$) then a new  channel of land is created  and the compatibility with the pattern $\wp$ will be preserved intact (no highway has been severed and no new causeway has been created). Viceversa if we split the double zero, necessarily we will create a new branchcut: at this point the reader does  not know yet if the newly created cut would be a causeway (which would be admissible) or an ``inverted'' causeway, which would violate admissibility. While we anticipate that the splitting would result in an inadmissible Boutroux curve, even in the other case we should not have performed the split because the newly created causeway would carry no traffic (since there was no highway prior to emersion) and hence violate the compatibility with the pattern $\wp$. The way the splitting occurs will be detailed in Sect. \ref{sectsing}.

Viceversa if {\bf (c0)} is the result of a sinking ($3\mapsto 0$)  then we may have to split the double root because a channel of emerged land is being flooded. This depends on whether this channel was traversed by a highway or not: if no highway was traversing it then the sinking of the saddle below waterlevel would not violate admissibility and also preserve the compatibility with the pattern $\wp$ ($3\mapsto 0 \mapsto 2$); if there was a highway then we are forced to split the saddle into two branchpoints ($3\mapsto 0 \mapsto 1$). 

%

By ``reversing the arrow of time'' we also see that if {\bf (c0)} is the result of the collapse of a causeway, the resulting saddle will emerge rather than sink ($1\mapsto 0 \mapsto 3$) and so preserve the connectivity.

\paragraph{Case {\bf (c2)}}
 \begin{wrapfigure}{l}{0.35\textwidth}
\begin{center}
\resizebox{4.5cm}{!}{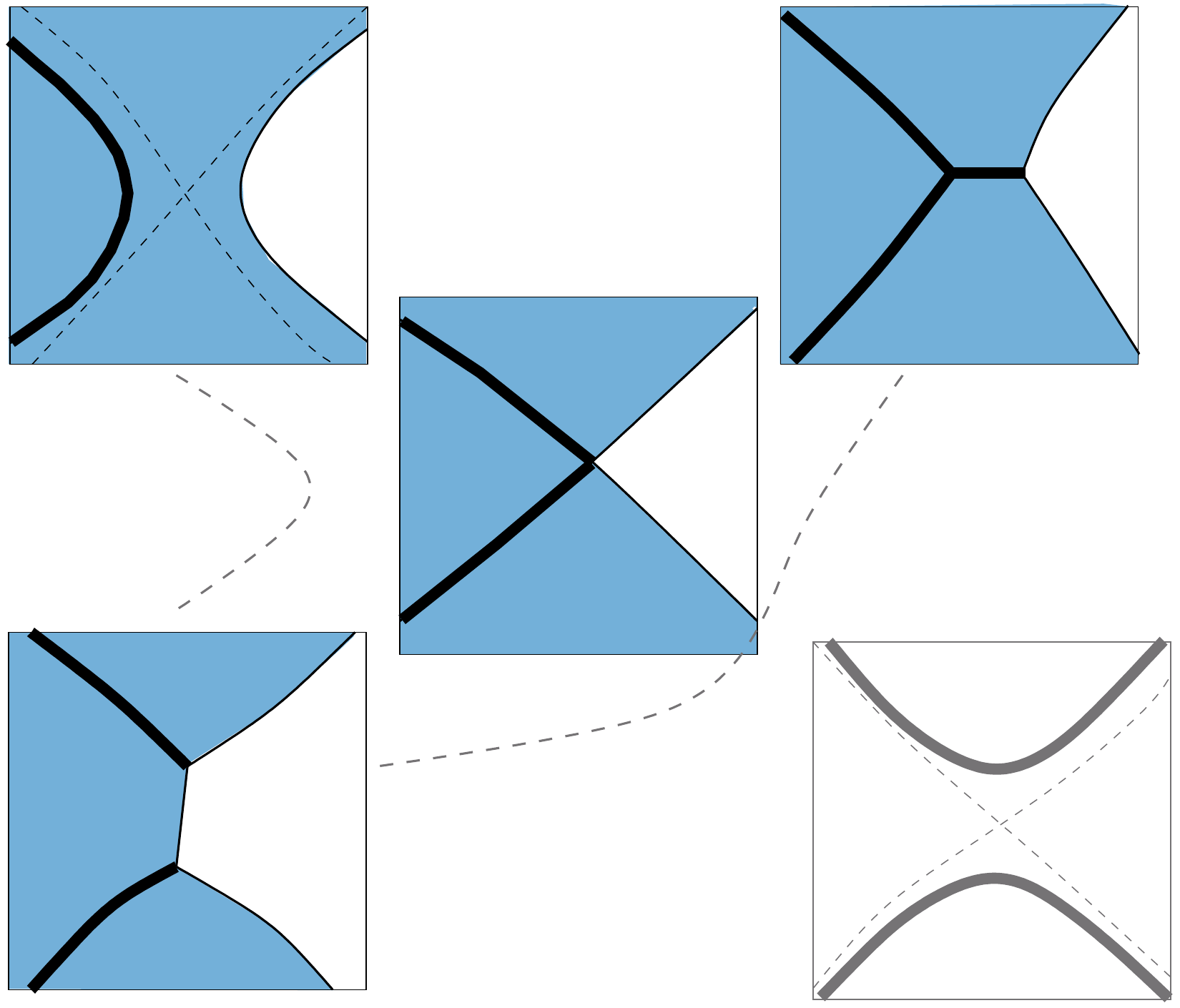}
\end{center}
\end{wrapfigure}
Can be the result of either merger of two branchpoints connected by a shoreline  ($1\mapsto 0$) or a causeway ($2\mapsto 0$) or emersion of a saddle through a causeway ($3\mapsto 0$)  but not of a sinking because the causeway  was surrounded by waters (due to admissibility) and hence the nearing saddle cannot but have emerged.
 If this is resulting from a collapsing shoreline, then, depending on the connectivity pattern, we may have to split again the double root  ($1\mapsto 0 \mapsto 2$) (the branchpoints ``scattering'' at right angles) or let the double root move away ($1\mapsto 0 \mapsto 3$) (in a sinking motion as will be proved later).

If this resulted from a collapsing causeway, then we are forced to split again the root ($2\mapsto 0 \mapsto 1$)  otherwise the saddle would ``emerge'' and severe the causeway (and also violate admissibility).

\begin{wrapfigure}{r}{0.35\textwidth}
\vspace{-1.5cm}
\begin{center}
\resizebox{4.5cm}{!}{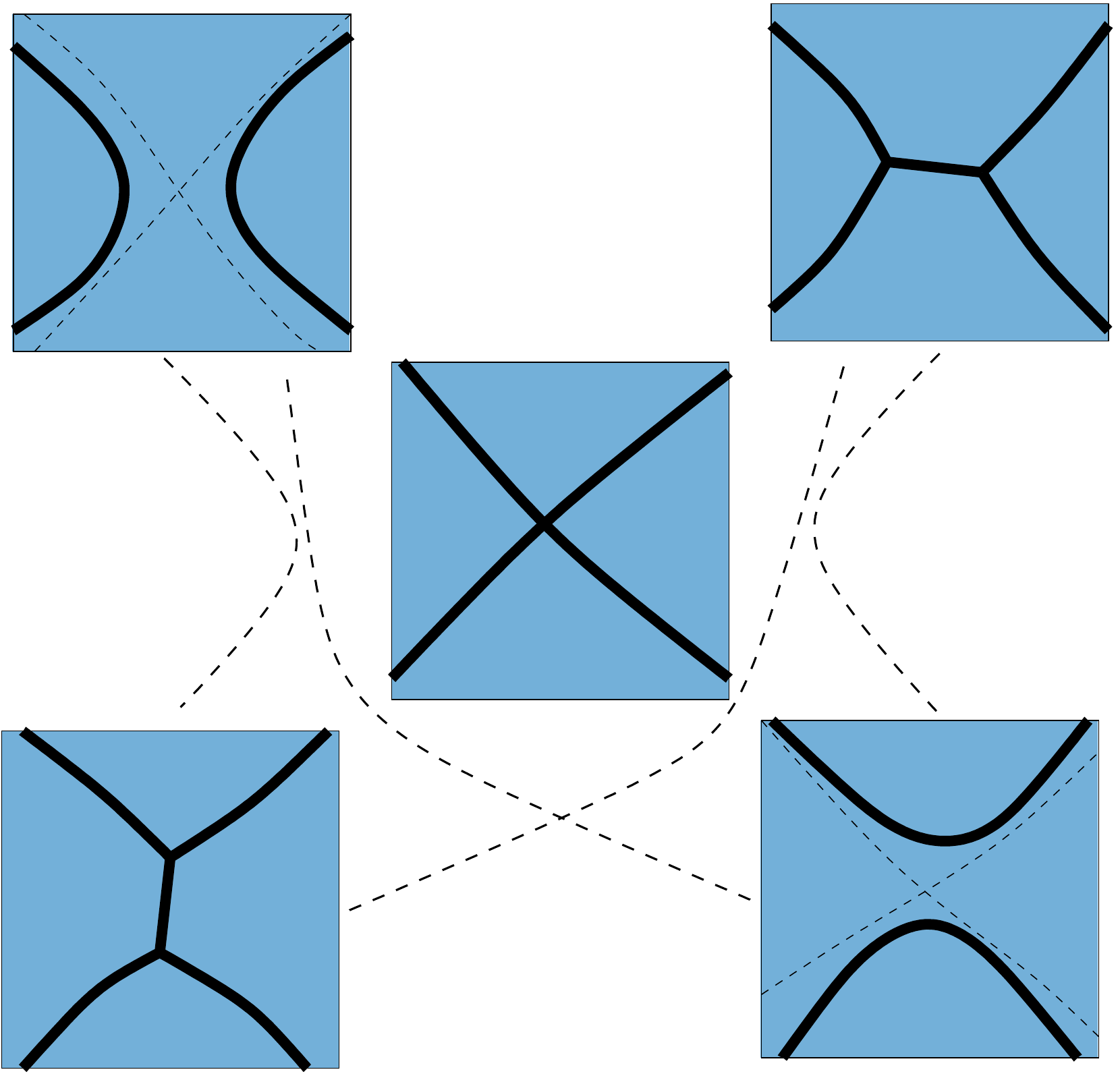}
\end{center}
\vspace {-2cm}
\end{wrapfigure}

\paragraph{Case {\bf (c4)}}
This can occur as the result of an emersion of a saddle through two opposite cuts ($4\mapsto 0$ or $3\mapsto 0$)  or the collapse of a causeway ($1\mapsto 0$ or $2\mapsto 0 $). In the former case we need to split the root and check that the split maintains compatibility with the pattern $\wp$; indeed if the root would split in the ``wrong way'' (i.e.  $4\mapsto 0 \mapsto 2$ or $3\mapsto 0 \mapsto 1$, but it does not) then the newly created causeway would carry no traffic, while the admissibility is not at stake. The splitting occurs in either ($4\mapsto 0\mapsto 1$) or ($3\mapsto 0 \mapsto 2$). The important point is that we cannot have ($3\mapsto 0 \mapsto 1$) or ($4\mapsto 0 \mapsto 2$)  because (as will be seen in Sec. \ref{sectsing}) this would imply a discontinuity in $(\dot V_t, \dot T_t)$ and precisely a change of sign in the tangent vector; as long as the path is {\em transversal} (in the sense specified in (\ref{transversality})) to the discriminant (and with nonzero tangent vector) then the ones indicated are the only possibilities.

\begin{wrapfigure}{r}{0.35\textwidth}
\begin{tabular}{cc}
 \includemovie[
   autoplay, autostop,repeat,toolbar,rate=2
 ]{2.5cm}{2.5cm}{Genus1-2deg4.avi}
 &\includemovie[
   autoplay, autostop,repeat,toolbar,rate=2
 ]{2.5cm}{2.5cm}{Genus1deg4.avi}\\[20pt]
 \includemovie[
   autoplay, autostop,repeat,toolbar,rate=2
 ]{2.5cm}{2.5cm}{Genus2-1deg4.avi}
 &\includemovie[
   autoplay, autostop,repeat,toolbar,rate=2
 ]{2.5cm}{2.5cm}{Genus2deg4.avi}
 \end{tabular}
 \caption{These animations show the process exemplified in the diagram. Specifically they correspond to the potential $V(x) = -\frac{x^4}4 +\frac C 2 x^2$ and $T\equiv 1$. The parameter $C$ goes from $-1$ to $1$ and the transition occurs at $C=0$.}
 \end{wrapfigure}

In the latter case we can  move the double root away or split it again (scattering the roots); this depends on the traffics that flow through the causeways. For example if we are in ($1\mapsto 0$) then the choice of going along ($1\mapsto  0\mapsto 4$) or ($1 \mapsto 0 \mapsto 2$) depends on the values of the traffics traversing the causeways in picture ($1$): if the traffic traversing the short causeway in ($2$) would be zero as a result then  this implies that we can and must evolve into ($4$).

\paragraph{Summary}
We see that this exhaustion list of possible scenarios shows that we can ``move away'' from a  critical situation in all cases by choosing an appropriate process that {\bf (a)} preserves the admissibility and {\bf (b)} the compatibility with the pattern of connectivity $\wp$.

The keen reader may have noticed that in fact the pictures are all the same, only the coloring changes, namely the position of the branchcuts and the definition of the height function $h$, which has to be continuous away from the critical point during the transition.

The list above relies on the precise way two cells of admissible Boutroux curves can be glued together along their boundaries; for example it is not possible to glue the cells corresponding to $(1)$ and $(2)$ in  case {\bf (c0)} as this would result in a manifold singular along the critical locus and prevent us from lifting the path $(V_t,T_t)$ across the discriminant, but rather it would only allow paths that ``bounce off'' the discriminant in the direction they came.

The next section is aimed at the precise analysis of the split/merge and how to glue cells. After that part is proven, Thm. \ref{mainthm} will be as well.

\subsection{Gluing cells along the boundary}
\label{sectsing}
In order to complete the proof of Thm. \ref{mainthm} after the picturesque discussion above we need to explain how 
to glue two cells of admissible Boutroux curves together to form a locally smooth manifold in which we can lift any path $(V_t,T_t)$ crossing the discriminant. 

Since we are lifting a curve, we are essentially solving an ODE in each cell, integrating the tangent vector 
\be
\dot \omega = \Pot_\star^{-1}(\dot V, \dot T)\ .
\ee
Most of the work has already been done in Section \ref{secdim} since we have completed each cell to a smooth manifold with boundary; the boundary points consist of the simply critical, simple, admissible Boutroux curves.
The whole point is the proof of the following

\bt
\label{thm_match}
Suppose that the lift  $y_t^2 = P_t(x)$  of a  $\mathcal C^1$  path $(V_t,T_t)$ in the space of admissible simple Boutroux curves (Lemma \ref{lemmalift})  is such that at $t=t_0$ one double zero $m_0$ is on the sea-level $\mathfrak X$ and is the limit of a double zero $m(t)\to m_0$.

Suppose also that  $0\neq \dot h(m)\big|_{t=t_0}\in \R$, namely the path $(V_t,T_t)$ intersects the the discriminant transversally (or --equivalently-- the saddle corresponding to $m(t)$ was either sinking or emerging with nonzero speed).

 Then we can continue the family of Boutroux admissible simple curves  $ y_t^2 = P_t(x) $ for $t\geq t_0$  in such a way that 
 \begin{enumerate}
\item $\Pot[y_t] = (V_t,T_t)$ is smooth at  $t_0$;
\item the root $m(t_0)$ splits into two simple roots $m_\pm(t)$. If $\dot h(m)\big|_{t=t_0} >0$ (the saddle was {\bf emerging}) then the two roots will split  along the opposite directions  of {\em steepest ascent}   of the saddle, as determined by continuity from the directions of steepest ascent for $t<t_0$. Viceversa if $\dot h(m)\big|_{t=t_0} >0$ (the saddle was {\bf sinking}) the roots will split along the steepest descent directions.
\end{enumerate}
The resulting family is $\mathcal C^1$ (but not $\mathcal C^2$) in the differential structure induced by $\mathbf P$.
\et



Before entering into the proof we want to put the theorem into context: by Lemma \ref{lemmalift}, which applies to simple curves in general, we {\em could} continue the lift of $(V_t,T_t)$  with a  family $y_t^2 = P_t(x)$ of Boutroux curves preserving the multiplicity of the critical saddle. However the discussion of the previous section (Sect. \ref{sectsimple}) has shown that we need to be able to ``decide'' whether a critical double root has to split or not, in order to preserve the admissibility or  the connectivity. 
The present theorem serves precisely this purpose by  showing that we can $\mathcal C^1$--glue a family $\{P_t\}_{t\leq t_0}$ of simple $\B$--curves with a moving saddle to a family $\{P_t\}_{t\geq t_0}$ of simple $\B$--curves with a {\em splitting} saddle; the gluing is completely ``transparent'' to the map $\Pot$, namely the path $(V_t,T_t)$ can be as smooth as desired.
 Moreover the theorem relates the directions of the splitting to the motion (ascending or descending) of the saddle prior to the  transition. 

{\bf Proof.}
Suppose that $P_t$ is the lift of the path $(V_t,T_t)$ in a cell $\mathfrak C_0$ of $\B_{reg}^{adm}$ such that the critical curve is the result of a double root (a saddle--point of $h(x)$) intersecting the sea-level, either by emersion or immersion.

Let $\dot \omega = \Pot_\star^{-1} (\dot V,\dot T)$ be the corresponding deformation of the Boutroux curve $y^2 = P_t(x)$, namely $\dot \omega = \frac {\dot P_t(x)\d x}{y_t}$. At time $t=t_0$ we have $h(m(t_0))=0$ and 
\be
\rho:= \dot h(m_{t_0}) = \Re \le(\int_{\a_{cr}}^{m_{cr}} \dot \omega_{cr}\ri)
\ee
 where $\rho\geq 0$ if the saddle was emerging from the sea or $\rho\leq 0$ if it was sinking: the assumption of {\em transversality} (Def. \ref{transversality}) of the path $(V_t,T_t)$ to the discriminant is tantamount requiring that $\rho \neq 0$.

At $t=t_0$ we have a simply critical Boutroux curve $y_{cr}^2 = P_{cr}(x)$ and the tangent vector $\dot \omega_{cr}$ points towards the exterior of the cell; therefore we must glue to it one of the cells that have the same boundary in which the same tangent vector $\dot \omega$ points {\em inside}.

There are three other cells of $\B_{reg}$ in a neighborhood of $[y_{cr}]$; one of them $\mathfrak C_1$ consists of simple Boutroux curves with a saddle just  above sea-level (if $\rho>0$, i.e. the saddle was emerging or viceversa below sea--level if $\rho<0$) and the other two $\mathfrak C_{2,3}$ consist of simple Boutroux curves with two branchpoints $m_\pm$ in a neighborhood of $m_{cr}$. 

It is obvious that we can glue our starting cell with $\mathfrak C_1$ since the vector $\dot \omega$ which ``lifts'' the saddle points (by definition) in the interior of the cell.

Of the other two cells one can be glued to $\mathfrak C_0$ and the other one to $\mathfrak C_1$, and we must establish which is which. From Sect. \ref{secdim} we know that any vector field of class at least $\mathcal C^0$ can by extended to the boundary of $\mathfrak C_{2,3}$; therefore we can integrate the ODE $\dot \omega = \Pot_\star^{-1}(\dot V,\dot T)$ in either cells. 

From Prop. \ref{propbingo} we know that the cell we need is such that the vector field is of the form 
\bea
\dot \omega = \dot \omega_{cr}  + \le|\frac 1{\ln|m_+(t)-m_-(t)|} \ri| \frac {S_1(x)\d x}{w_t} + \dots\cr
\rho =  \res{x=m_{cr}} \frac {S_1(x)\d x}{w_{cr}} \in \R
\eea
\begin{wrapfigure}{r}{0.35\textwidth}
\vspace{-1cm}
\begin{center}
\resizebox{5cm}{!}{
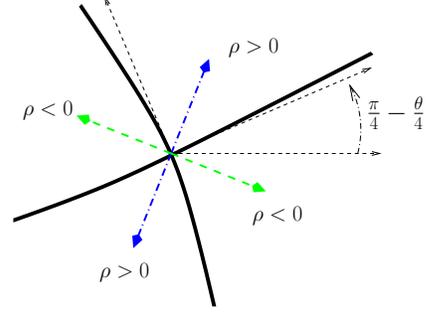}
\end{center}
\caption{
From the double zero $m_{cr}=0$ originate four critical trajectories that  depart along directions $\frac {\pi-\theta}4 + k\pi/2 $,
where   $\theta = \arg M^2_{cr}(0)R_{cr}(0)$.
The double root  (a saddle point of $h(z)$) then splits into two simple zeroes $m_\pm$  that come out along the directions that bisect the angles formed by the critical trajectories. For $\rho >0 $ or $\rho<0$  the split occurs along the indicated directions. Since $M_{cr}(0)\sqrt{R_{cr}(0)}$ is the Hessian of $h_{cr}$, one can check that the directions indicated are also the steepest ascent (for $\rho>0$) or descent ($\rho<0$) directions at the saddle.}
\label{Figsplit}
\vspace{0cm}
\end{wrapfigure}
We are going to show that $S_1(m_{cr})\neq 0$ determines the directions in which the roots $m_\pm(t)$ split (or merge) according to the statement of the theorem. In addition we will determine the order in $\delta t = t-t_0$ of the split.
For simplicity (without loss of generality) we assume $m_{cr} =0$; 
using Prop. \ref{propbingo} for $\dot \omega $ and hence $\dot P_t = M(x) \dot \omega/\d x$ we have (setting $\epsilon = |m_+- m_-|$)
\bea
P_t(x) \sim  P_{cr}(x) + \delta t M_{cr} (x)\le( S_{cr}(x) + \frac 1{|\ln\epsilon |} S_1(x)  \ri) =\cr
x^2 M^2_{cr}(x) R_{cr}(x) + \delta t x M_{cr}(x)\dot R(x) + \frac {\delta t}{|\ln( \epsilon )|}M_{cr}(x)  S_1(x)
\eea
where $w_{cr}^2 = x^2 R_{cr}(x)$, $R_{cr}(0)\neq 0$ and $S_1(0)$ is given by eq. (\ref{bingo}) and we have used that $S_{cr}(x)$ vanishes at $m_{cr}=0$. This implies that the roots $m_\pm(t)$ behave as 
\bea
m_\pm  = \pm \sqrt{\frac   {\delta t}{|\ln( \epsilon )|}    }\sqrt{\frac {S_1(0)}{M_{cr}(0)R_{cr}(0)}}(1+o(1)) = \cr
=\pm \sqrt{\frac   {\delta t}{|\ln( \epsilon )|}    }\sqrt{\frac \rho {M_{cr}(0) \sqrt{R_{cr}(0)}}}  (1+o(1))
\eea
From this we find 
\be
\epsilon^2 \ln \epsilon  = \delta t \,\frac \rho {M_{cr}(0) \sqrt{R_{cr}(0)}}(1+o(1))\ \ \ \Rightarrow \ \ \ 
\ln \epsilon  = \ln \delta t(1+o(1))
\ee
and thus, finally,  
\be
m_\pm = \pm \dot m \sqrt{\frac{\delta t}{\ln \delta t}} (1 + o(1)) \ \ 
,\ \ 
\dot m ^2 = \frac{\rho}{M_{cr}(0) \sqrt{R_{cr}(0)}}
\ee

The determination of $\sqrt{R_{cr}(0)}$ is the same used in the definition of the height function $h(x)$ by continuity from before the critical situation, when the saddle was not on a branchcut (and hence no ambiguity of sign exists). If a harmonic function is written locally $H(x)= \Re (c x^2)+ \dots$ then the direction of steepest ascent are $-\arg(c)/2 +k\pi$ whereas those of steepest descent are $-\arg(c)/2 + \pi/2 + k\pi$. In our case $h(x) \sim \Re M_{cr}(0) \sqrt{R_{cr}(0)} x^2$ and hence if $\rho>0$, $\pm \dot m$ point in the steepest ascent directions as depicted in Fig. \ref{Figsplit}.

The last statement about the smoothness class is immediate since we have seen that  for some polynomials $
\dot P, \ \dot R$, 
\be
P_t = P_{cr} + \delta t \dot P + \frac {\delta t}{\ln \delta t} \dot R + \dots
\ee
and hence the {\em coefficients} of $P_t$ is not $\mathcal C^2$ at $\delta t=0$. Note, however that $\Pot[y_t]$ is as smooth as $(\dot V, \dot T)$ is. 
{\bf Q.E.D.}\par
\vskip 30pt

\subsection{Uniqueness of the Boutroux curve compatible with the connectivity pattern}
\label{secunique}
The process by which we have constructed our final Boutroux curve with external potential $V$ and total charge $T$ does not prove uniqueness since the discriminant of simply critical simple curves  $\mathbf \Sigma_{reg}^0$ is probably not even connected;  indeed the singular discriminant (corresponding to critical non-simple Boutroux curves) has codimension $1$ inside the regular discriminant and thus a homotopy argument seems difficult to implement unless one analyzes in more detail the gluing of the cells of the regular discriminant that meet at generic points of the singular discriminant.

The simplest workaround to this problem is actually that of invoking the asymptotic theory of pseudo--orthogonal polynomials. Indeed in \cite{BertoMo} it was shown that any admissible Boutroux curve can be used to construct the $g$--function for the asymptotic study of the OPs that correspond to the potential of the curve, for a certain connectivity pattern. If there were more than one admissible Boutroux curve compatible with the same connectivity pattern $\wp$ then we would have the contradicting situation that the same polynomials have different strong asymptotics for  large degrees (or on suitable subsequences). This  contradiction proves uniqueness, although it seems like a far-fetched solution and probably a more direct proof can be found using some refined homotopy argument.

\br
It would be interesting to count the degree of $\Pot$. For example we claim (without proof) that if $V(x)$ has simple critical points (and distinct critical values of $\Re V(x)$) then for small $T$ there are $d$ admissible Boutroux curves in $\Pot^{-1}(V,T)\cap \B^{adm}$. They correspond to the deformation of the nodal curve $y^2 = (V'(x))^2$ obtained by splitting one of the roots of $V'(x)$ (which is a node of $y$). The splitting of the critical point with lowest critical value of $\Re V(x)$  is compatible with any connectivity pattern where the opened causeway carries some traffic. All sectors are connectible to all other sectors, but only one is ``essentially'' connected and it is the one on whose shoreline there is an entry-point to the causeway.  If that sector is not  accessed in our connectivity pattern, then a different critical point must be used.

Note that the degree of $\Pot:\B^{adm}\to \V$ is {\bf not constant} (only locally constant) since --as we have seen-- the number of cells merging at the regular discriminant is not always the same (either three or four cells merge at a point of $\Delta_{reg}^0$). The degree of $\Pot:\B\to \V$ instead is well defined (always four cells merge at the discriminant). 
\er

\subsubsection{Summary of the proof of Thm. \ref{maintheorem}}
Since Thm. \ref{maintheorem} was stated quite far we recall what constitutes the proof of Thm. \ref{maintheorem}.

In Section \ref{secbasecurve} we have shown that we can construct a simple noncritical admissible Boutroux curve compatible with any given connectivity pattern $\wp$. 

Next, in Section \ref{secdeform} we have shown that (Thm. \ref{mainthm})  that from a given simple admissible noncritical Boutroux curve with a connectivity pattern $\wp$ we can smoothly deform it so that the potential and charge match any chosen (generic) one and along the deformation we can preserve the same connectivity pattern by choosing appropriately the branch of the inverse map $\Pot^{-1}$ near the phase transition that may occur.

Finally in Section \ref{secunique} we have proved the uniqueness of the result by invoking the uniqueness (along suitable subsequences) of the strong asymptotic of the corresponding generalized orthogonal polynomials.

\section{Conclusion and final comments}
The aim of the paper was to prove the existence of  a suitable $g$--function for the implementation of the nonlinear steepest descent method. The requirements for the $g$--functions have been translated into a set of algebraic/harmonic requirements and these latter have been analysed. 

As pointed out in the course of the proof some requirements are specific to the 
case of  the pseudo--orthogonal polynomial,  but in fact  the techniques can be extended to cover similar situation like 
\begin{itemize}
\item Painlev\'e\ equations and higher generalizations; in this case  we need to allow for Boutroux curves where a branchpoint may be at infinity (i.e. of odd degree) and the notion of admissibility should be relaxed/modified keeping into account the triangularity of the Stokes matrices;
\item pseudo--orthogonal polynomials with hard-edges and with  potentials with rational derivatives \cite{BEHsemi};
\item Laurent biorthogonal polynomials (including OP on the circle) for log-rational symbols  with hard edges \cite{BertoMisha}.
\end{itemize}
The scope of the discussion  of how degenerations occur needs to be broadened to cover non--simple situations,  for example in the case relevant to $2$--D gravity \cite{ItsKitaev, ArnoDuits}. In this case, due to the reality constraint on the external potential, the generic degeneration actually involves a double root coalescing with a branchpoint and then this triple branchpoint splitting three-ways. It can be seen that there are {\bf two ways} in which this triple splitting may occur: one in which one root moves away analytically and the other two as square-roots of $(t-t_0)$, and the other in which the three roots move away according to third roots of $t-t_0$.

For higher order coalescence the ``zoology'' of the splitting becomes immediately quite large; the issue is interesting from the point of view of the geometry of the moduli space of  Boutroux curves (study of the local properties of the discriminant near the singularities) and also for the application to Riemann--Hilbert analysis.

Indeed the study of the parametrix for the pseudo orthogonal polynomials near a non-simple turning point (or order $2,3,4,\dots$)  requires the use of some other Painlev\'e\ like transcendents. For example in \cite{ArnoDuits} the turning point of order $3$ has five critical trajectories and the local parametrix is constructed in terms of the $\psi$--function for Painlev\'e\ I. In general of those critical trajectories some or all may be branchcuts and so the parametrix needs to be ``patched'' from pieces like in the Airy case \cite{BertoMo} corresponding to a simple turning point but with three cuts and not only one.
Additionally some of the main interests are also in ``double scaling'' approaches in which the splitting (or coalescing) is scaled together with the large parameter: for example in the recent \cite{eynardcut} it is studied the ``birth of a cut'' in a double scaling perspective  although without explicit use of Riemann--Hilbert techniques (although it could be addressed using parabolic cylinder functions i.e. Hermite functions, which appear already in the paper on physical grounds). 
In general, for a double scaling study of higher order degenerations in the local parametrix will contain its own large parameter and hence its own $g$--function.
\subsection{Relation to the quasi--linear Stokes' phenomenon of Painlev\'e\ II}
\begin{wrapfigure}{r}{0.28\textwidth}
 \begin{center}
 \vspace{-0.5cm}
 \includemovie[
   autoplay, autostop,repeat,toolbar,rate=2
 ]{3.7cm}{3.7cm}{PainleveII.avi}
 \end{center}
\end{wrapfigure}
To conclude these remarks we would like to point out the relevance of the phase transitions described in this text to the so--called {\em quasi--linear Stokes' phenomenon} (\cite{ItsBook} and references therein).
In the case of PII, for large modulus of the argument $\xi$ of the Painlev\'e\ transcendent $\eta$ (we do not use the usual symbols $x,y$ here because we already used them for other purposes in the text) the asymptotics of $\eta$ is expressible in terms of elliptic functions evaluated at some point  $\Xi$ in the Jacobian. The elliptic curve is precisely the spectral curve we consider in this text for the potential 
\be
V_{PII}(x) = i\le(\frac {4}3 x^3 + {\rm e}^{i\varphi} x\ri)\ ,\ \ \ T_{PII} =0\ .
\ee
where $\varphi = \arg(\xi)$.

\begin{wrapfigure}{r}{0.3\textwidth}
\includegraphics[scale=0.24]{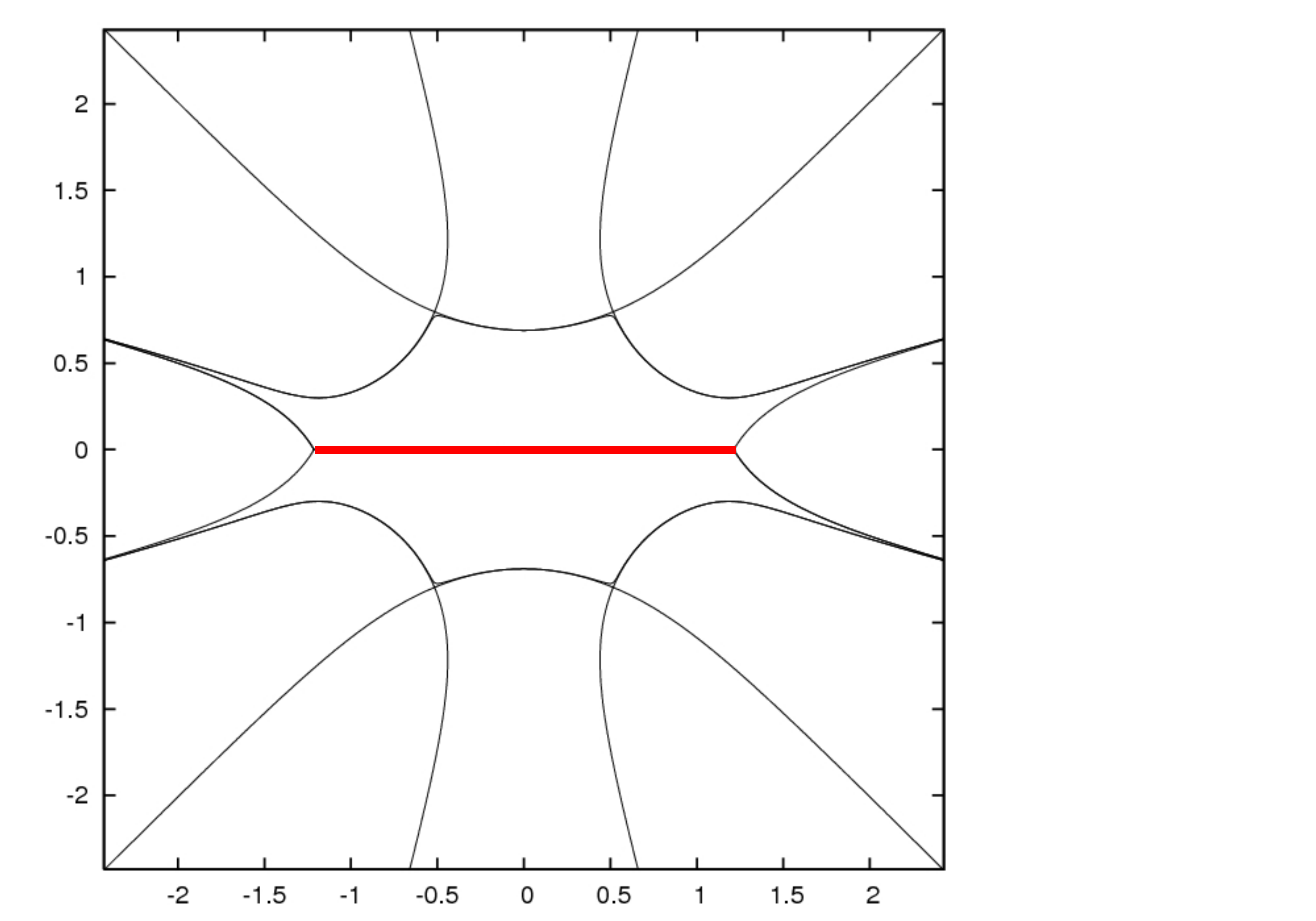}
\caption{An admissible  Boutroux curve  of genus $0$ for the potential $V = \frac {x^6}6$ and total charge $T=1$. The cut is marked in red.}
\label{x6genus0}
 \end{wrapfigure}

The notion of admissibility needs to be modified (the Stokes matrices of PII are not all upper-triangular) from the one used in this paper, but the general philosophy remains unaltered. In particular the Boutroux curve becomes critical when $\varphi$ is on the {\em canonical rays} and the genus drops from $1$ to $0$. This is precisely one of our ``phase transitions'' for the particular path in $\V$ space parametrized by $\varphi\in [0,2\pi)$ (see animation). The dependence of the asymptotics of $\eta$ on $\varphi$ and its sudden ``discontinuous'' changes are what underlies the quasi--linear Stokes' phenomenon.

In our setting the r\^ole of the Painlev\'e\ transcendent is played by  the isomonodromic tau function of the system solved by the (generalized) orthogonal polynomials \cite{BEHsemi} and the parameter $N$ (the degree of the OPs) plays the r\^ole of the modulus of $\xi$ in PII: since isomonodromic tau functions enjoy the Painlev\'e\ property, the sudden changes of genus and/or topology of the critical graph of our Boutroux curve w.r.t. the potential/charge, are the direct analogue of the quasi--linear Stokes' phenomenon described above for PII.
We also mention that the q.l. Stokes' phenomenon depends also on the Stokes' parameters of the associated linear ODE: these are --in our present description-- the {\em traffics}. Indeed changing the traffics for a given connectivity pattern may result in a highway to carry no net traffic, and this would  also induce a sudden change in the connectivity of the Boutroux curve.

{\bf \large Acknowledgments}\\[4pt]
The author would like to thank A. Its, P. Bleher, P. Wiegmann and S. Y. Lee for discussions related to the topics and possible applications.

\appendix
\section{Some examples}
\begin{figure}
\vspace{-1cm}
\includegraphics[scale=0.24]{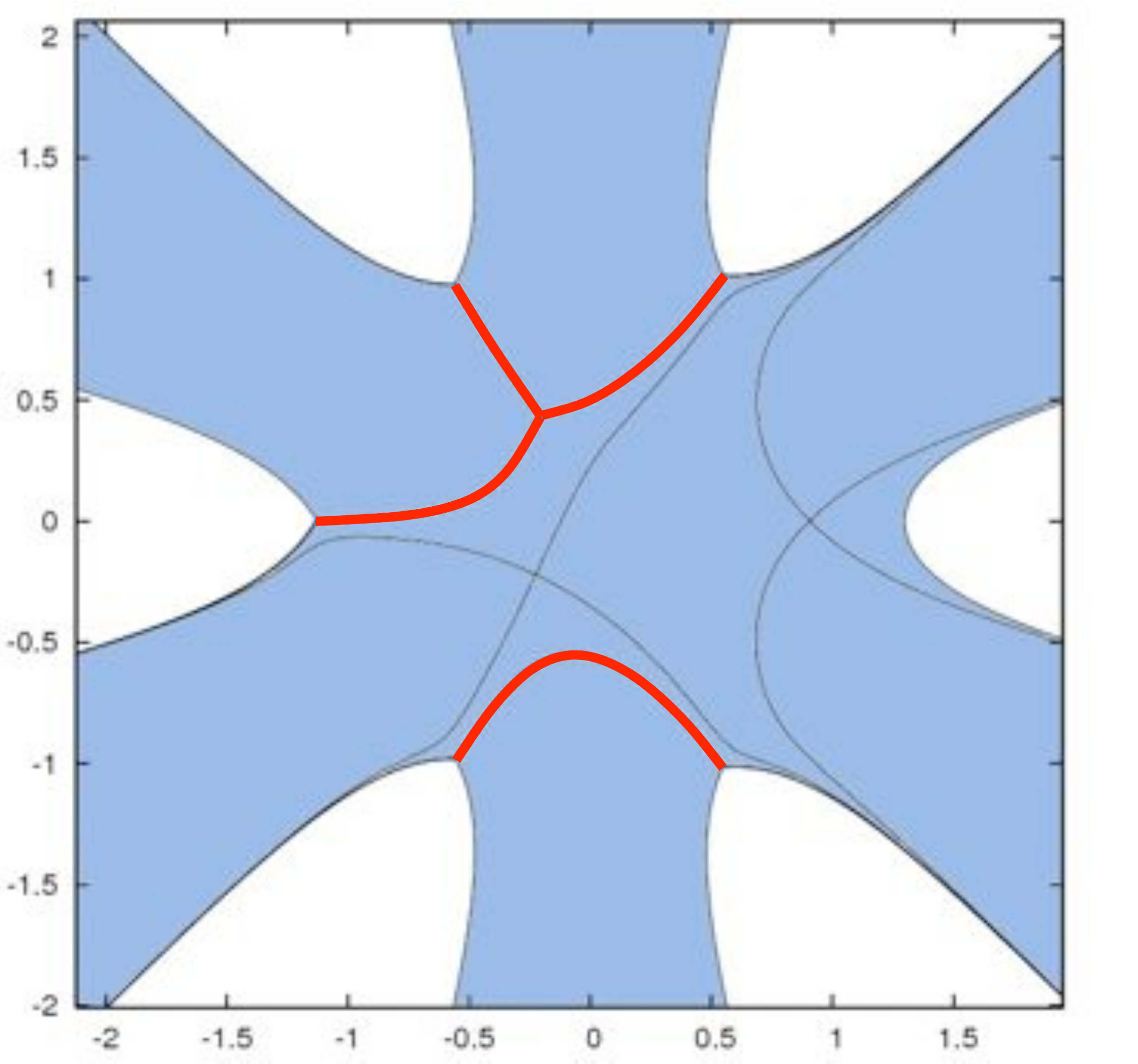}
\includegraphics[scale=0.24]{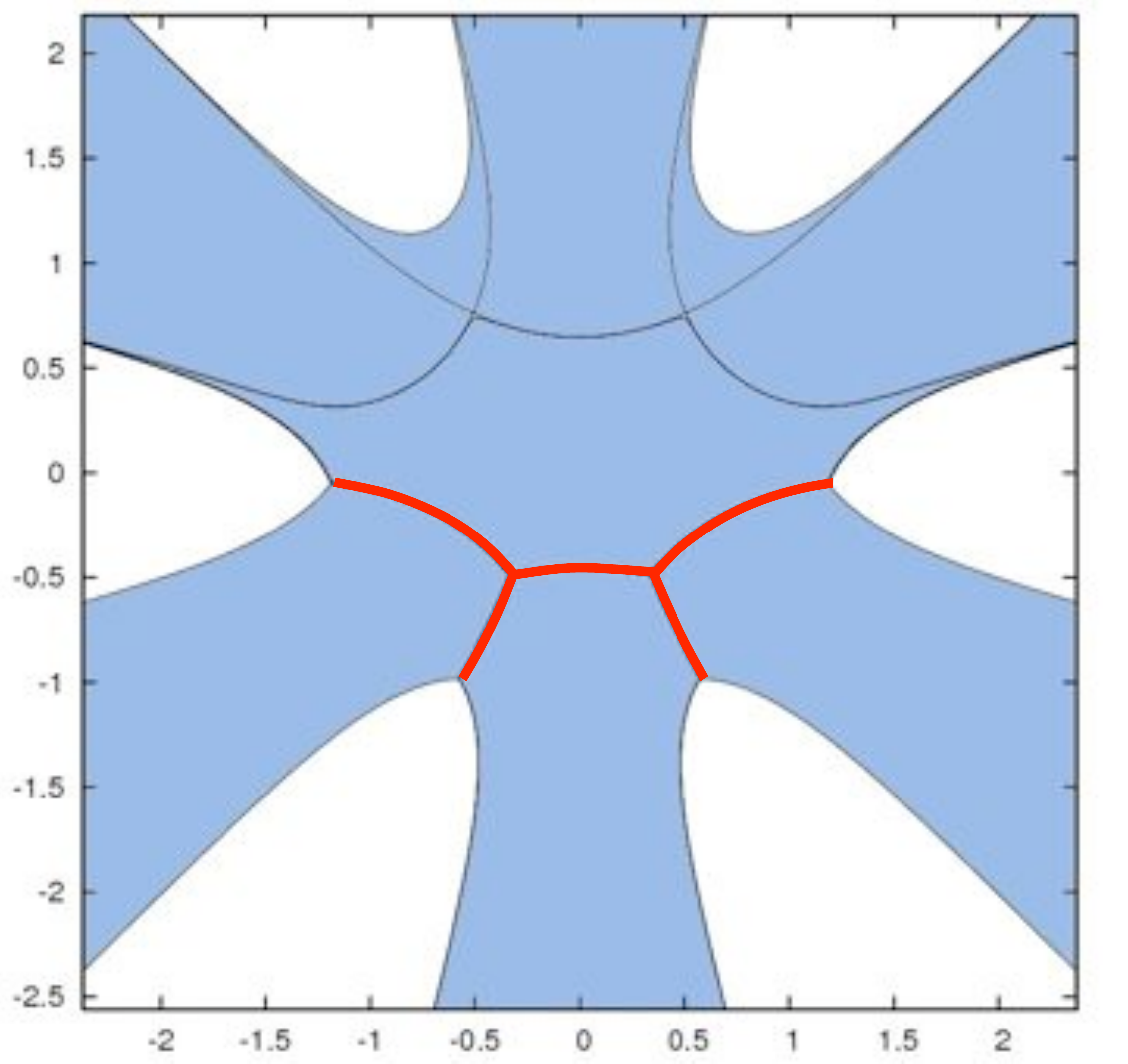}
\includegraphics[scale=0.24]{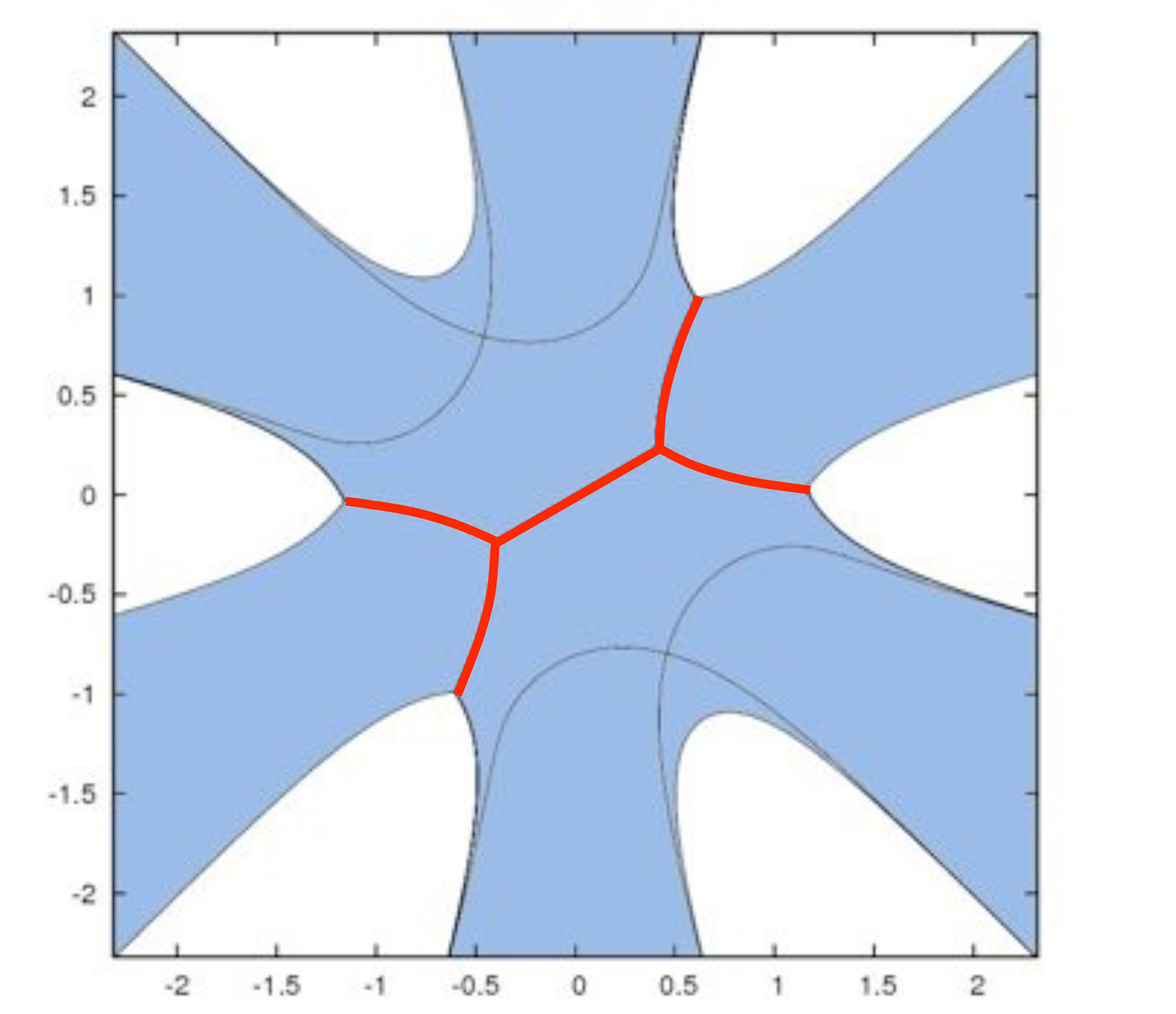}\hspace{-1cm}

\caption{Several admissible  Boutroux curves  of genus $2$ for the same potential $V = \frac {x^6}6$ and same total charge $T=1$, but compatible with  different connectivity patterns. The cuts are marked in red. The first one is the same curve depicted also in Fig. \ref{3Dgalore}.}
\label{Fig9}
\end{figure}

It is worth mentioning that the same point $(V,T)\in \V$ may correspond to a non-simple Boutroux curve for a certain connectivity pattern, while for a different connectivity pattern corresponds to a simple one: in other words the discriminant is only on one ``sheet'' of the map $\Pot$.

A simple example of this situation is the potential $V(x)= x^6/6$; if we require the maximal connectivity pattern (i.e. all sectors are joined to all others) then the origin is a zero for the Boutroux curve of order $4$ (for any total charge $T$)  and then there are $6$ simple zeroes 
\be
y^2 = x^4 (x^6-2T).
\ee
This family is non-simple and (non simply) critical for all $T\in \R_+$.
However one can find a family of admissible Boutroux curves (the ``usual ones'': in Fig. \ref{x6genus0} for $T=1$) of genus $0$ which are simple  for $T\in \R_+$ and correspond to a connectivity where only the left and right sectors containing the real axis are connected (only one highway with traffic, namely the real axis)
\be
y^2 = (x^2-A)\le( x^4 + \frac A2 x^2 + \frac 3 8 A^2\ri)^2\ ,\ \ A = \frac 2 5 50^{\frac 1 3} T^{\frac 1 3 }
\ee

More examples (found numerically using the algorithm described in the next appendix) are the ones in Fig. \ref{Fig9} which have two saddle points.

\section{Numerical algorithm}
We would like to explain briefly how to find numerically some nontrivial examples of Boutroux curves (admissible and not). We will not enter in any software details and leave it to the keen reader to implement the algorithm. We have produced a certain number of Octave programs (an open-source program of numerical manipulations similar to other commercial ones and available for several platforms) based on the idea that we are about to explain: these are available upon request. The results are the pictures found in the present paper and the short animations to be found on the author's webpage.
We  claim (leaving the straightforward check to the reader) that  any algebraic curve of the following form (here $v_{d+1}$ is the leading coefficient of $V'(x) = v_{d+1} x^d + \dots$)
\be
y^2 = (V'(x))^2 - 2 T v_{d+1} x^{d-1} +Q_{d-2}(x) = P(x)
\ee
has the property that 
\be
y \sim V'(x) - \frac T x+ \mathcal O(x^{-2})\ .
\ee
The polynomial $Q(x) = Q_{d-2}(x)$ is an arbitrary polynomial of degree $d-2$.
For generic $Q$ the roots of $P$ are simple and the genus is $d-1$.

\begin{wrapfigure}{r}{0.3\textwidth}
\vspace{-1cm}
\includegraphics[scale=0.24]{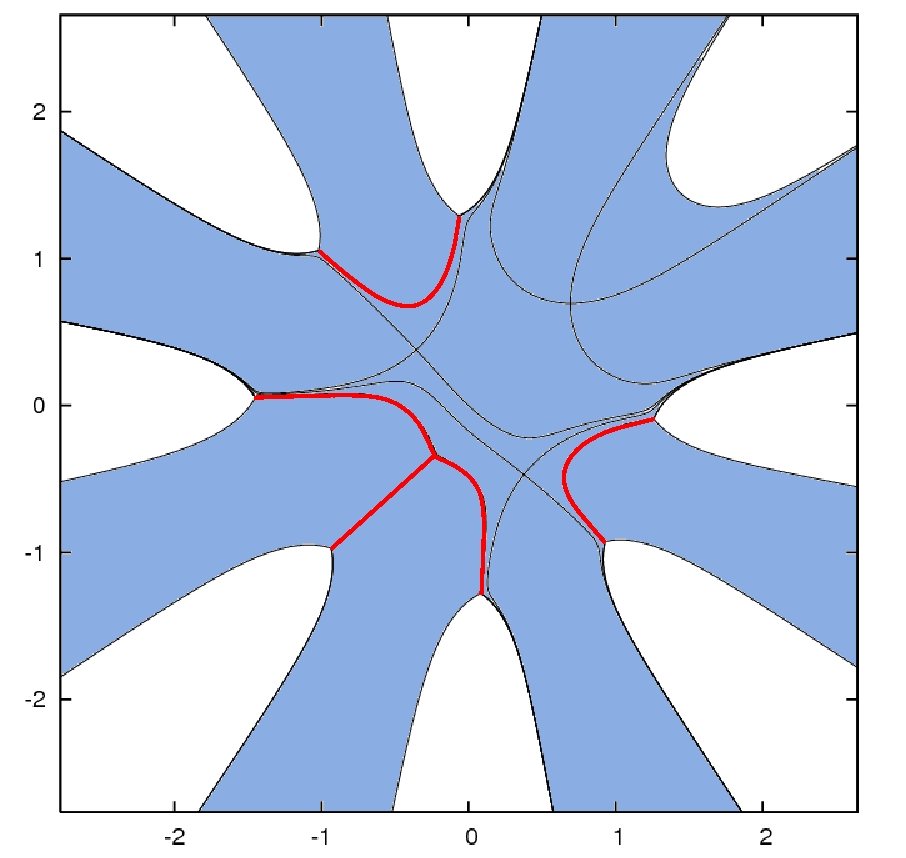}
\caption{An admissible Boutroux curve for the potential $V(x)= \frac {x^8}{4} + \frac {ix^6}6  -\frac {x^2}2  +3x$, $T=10$.}
\end{wrapfigure}

Choose randomly the coefficients of  $Q(x)$: we will show that we can define an ODE in $Q(x)$ that converges to some Boutroux curve for $(V,T)$.

We find the roots of $P(x)$ and choose a basis of cycles $\gamma_1,\dots, \gamma_{2g}$ spanning the homology of the curve. Define 
\be
\epsilon_j := \Re \oint_{\gamma_j} y \d x\ , \ \ \mathcal F := \frac 12\sum_{j=1}^{2g} \epsilon_j^2>0.
\ee 
The functional $\mathcal F$ vanishes if and only if the curve is Boutroux. We want to define an infinitesimal deformation of $Q$ that decreases $\mathcal F$. 

Indeed if we deform $Q$ by $\delta Q$ we have
\be
\delta \epsilon _j  =\Re \oint_{\gamma_j} \frac {\delta Q}{y} \d x 
\ee
Since the differential in the integral  above is the most general {\em holomorphic} differential ($\deg \delta Q\leq d-2 = g-1$), we can fix arbitrarily  the real parts of its periods; in particular the choice that maximizes the rate of decrease of $\mathcal F$ is in the opposite direction to  the  gradient of $\mathcal F$ with respect to the $\epsilon_j$'s, namely we need to find $\delta Q$ so that 
\be
\Re \oint_{\gamma_j} \frac {\delta Q}{y} \d x  = -\epsilon _j = - \Re \oint_{\gamma_j} {y} \d x\ , \ \ j=1,\dots, 2g\ .
\ee
By the general theory \cite{FarkasKra} such differential exists and is unique. The numerical integration (step-by-step) of this flow will produce eventually a Boutroux curve. Whether the result is admissible or not is decided by inspection of the final topology of critical trajectories.

Note that different choices of the homology basis $\gamma_j$ and/or the starting polynomial $Q$ may produce different final results: moreover the numerics (at least in the  implementation of the author, who is no expert in the field) becomes rapidly unstable (already degree $d=9/10$ pushes the limit of machine precision and one should then use some package with arbitrary precision arithmetics).

In order to obtain Boutroux curves with saddles we can require that $Q(x)$ be chosen so that $P(x)$ has a certain number of double zeroes at all times along the flow: this is easily implementable if the number of double points $k$ that we want to impose is less than $\le[\frac {d-1}2\ri]$. Indeed in this case we can parametrize $Q(x)$ instead than by its coefficients, by  $d-1$  zeroes  (counted with multiplicities) amongst the ones of $P(x) = (V'(x))^2 - 2T v_{d+1}x^{d-1} + Q(x)$. The positions of these zeroes are free parameters.

However if we want to look for ``low genus'' Boutroux curves, the algorithm becomes (numerically) unfeasible because these zeroes must belong to a certain algebraic variety describing some high order vanishing of the discriminant of $P(x)$.

\end{document}